\documentclass[a4paper,fleqn,usenatbib]{mnras}

\usepackage{newtxtext,newtxmath}
\usepackage{subfig}

\usepackage[T1]{fontenc}
\usepackage{ae,aecompl}

\usepackage{graphicx}	
\usepackage{amsmath}	
\usepackage{amssymb}	
\usepackage{cprotect}

\usepackage{etoolbox}
\makeatletter
\patchcmd\@combinedblfloats{\box\@outputbox}{\unvbox\@outputbox}{}{%
  \errmessage{\noexpand\@combinedblfloats could not be patched}%
}%
\makeatother

\title[Unveiling the nuclear radio structure of NGC\,6217]{Unveiling the 100\,pc scale nuclear radio structure of NGC\,6217 with e-MERLIN and the VLA}

\author[D.R.A. Williams et al.]{D.R.A. Williams,$^{1,2}$\thanks{E-mail: david.williams@physics.ox.ac.uk}
I.M. McHardy,$^{1}$
R.D. Baldi,$^{1}$
R.J. Beswick,$^{3}$
M. Pahari,$^{1}$
\newauthor{M.K. Argo,$^{3,4}$
A. Beri,$^{1,5}$
P. Boorman,$^{1}$
E. Brinks,$^{6}$
B.T. Dullo,$^{7,8,9}$
D.M. Fenech,$^{10}$}
\newauthor{J. Ineson,$^{1}$
P. Kharb,$^{11}$
J.H. Knapen,$^{8,9,12}$
T.W.B. Muxlow,$^{3}$
J. Westcott$^{6}$}
\\
$^{1}$School of Physics and Astronomy, University of Southampton, Southampton, SO17 1BJ, UK\\
$^{2}$Department of Physics, University of Oxford, Denys Wilkinson Building, Keble Road, Oxford OX1 3RH, UK\\
$^{3}$Jodrell Bank Centre for Astrophysics, School of Physics and Astronomy, The University of Manchester, Manchester, M13 9PL, UK\\
$^{4}$University of Central Lancashire, Jeremiah Horrocks Institute Preston, UK PR1 2HE\\
$^{5}$DST-INSPIRE Faculty, Indian Institute of Science Education and Research (IISER), Mohali, Punjab, 140306, India\\
$^{6}$Centre for Astrophysics Research, University of Hertfordshire, College Lane, Hatfield, AL10 9AB, UK\\
$^{7}$Departamento de  Astrof\'isica y Ciencias de la Atm\'osfera, Universidad Complutense de Madrid, E-28040 Madrid, Spain\\
$^{8}$Instituto de Astrof\'{i}sica de Canarias, V\'{i}a L\'{a}ctea S/N, E-38205 La Laguna, Spain\\
$^{9}$Departamento de Astrof\'{i}sica, Universidad de La Laguna, E-38206 La Laguna, Spain\\
$^{10}$Astrophysics group, Cavendish Laboratory, University of Cambridge, UK\\
$^{11}$National Centre for Radio Astrophysics - Tata Institute of Fundamental Research, Pune University Campus, Post Bag 3,\\Ganeshkhind, Pune 411007, India\\
$^{12}$Astrophysics Research Institute, Liverpool John Moores University, IC2, Liverpool Science Park, 146 Brownlow Hill, Liverpool,\\L3 5RF, UK\\
}
\date{Accepted XXX. Received YYY; in original form ZZZ}

\pubyear{2018}

\begin{document}
\label{firstpage}
\pagerange{\pageref{firstpage}--\pageref{lastpage}}
\maketitle

\begin{abstract}

We present high-sensitivity 1.51\,GHz e-MERLIN radio images of the nearby galaxy
NGC\,6217. We resolve the compact radio source at the centre of NGC\,6217 for the first
time, finding a twin-lobed structure with a total linear size of
$\sim$4 arcsec ($\sim$400\,pc). The radio source does not have a
compact central core but there is an unresolved hot spot near the
outer edge of the southern lobe. Combining our e-MERLIN data with new
VLA A-Array data and archival multi-wavelength 
data, we explore possible scenarios which might explain
this complex radio morphology. We conclude that this radio source is
most likely powered by a low-luminosity AGN (LLAGN) but with a
possible important contribution from nuclear star formation. We also
investigate the origin of a putative X-ray jet in NGC\,6217,
previously suggested in the literature. This `jet' consists of three
X-ray `knots' in a line, pointed away from the nucleus, with a total
size of $\sim$3 arcmin ($\sim$15\,kpc). We find no extended radio
  emission coincident with the `jet'. An additional compact radio
  source, aligned with the knots but without an X-ray counterpart is detected. We detect IR/optical sources falling
within the X-ray extraction regions of the `knots', but note that these sources could be chance associations due to high source density around the target field and we estimate the probability of three randomly aligned X-ray sources to be 0.3 \textit{per cent} in this field. 
\end{abstract}




\section{Introduction}
\label{sec:Introduction}
Active Galactic Nuclei (AGN), such as quasars and radio galaxies, can exhibit powerful collimated radio jets on scales of $\sim$100\,kpc \citep[e.g. ][]{Bagchi2007,Hocuk2010}. Some low-luminosity AGN (LLAGN) can host small, kpc-scale radio jets or bubbles \citep{Elmouttie1995,Irwin2003,Hota2006,Kharb2006,Kharb2010,Nyland2017,Capetti2017,Panessa2019} but when identifying these regions, caution must be taken to remove any contamination from star formation regions, supernovae and starburst super winds \citep{Pedlar1985,Baum1993,Heckman1993,Colbert1996,BaldiLeMMINGs,HoUlvestad,Ramirez2018,Panessa2019}. While the radio jets in AGN have been studied for decades, the presence of co-spatial X-ray emission is a more recent discovery \citep{Tavecchio2000,Schwartz2000,Chartas2000,Celotti2001,Sambruna2001,Harris2006, Worrall2009}. However, there are very few objects near enough to spatially resolve these X-ray jets, so new candidates for radio/X-ray/optical follow-up are useful; for example, 3C273 \citep{Sambruna2001, Jester2005, Harris2017}, M\,87 \citep{Marshall2002,Wilson2002} and Centaurus A \citep{Kraft2002,Hardcastle2003} are just three objects which can be resolved, but are all radio-loud sources, defined as the ratio of radio and optical luminosities, L$_{radio}$/L$_{optical}$ $>$ 10 \citep{Kellerman1989}, but see also e.g. \citet{Terashima,Cirasuolo2003,Capetti2010,Kharb2012,Bonzini2013,Padovani2017,Padovani2017b,Panessa2019}.

One potential object for resolving and studying low-powered X-ray jets is the barred spiral galaxy NGC\,6217. \citet[][hereafter PA01]{PietschArp} compared radio images of NGC\,6217 from the NVSS \citep{Condon87} with \textit{ROSAT} X-ray contour plots and claimed an extension south-west of the nucleus of NGC\,6217 that was aligned in both wavebands. PA01 hence concluded that it was possible that the X-rays could form the base of a radio jet. However, with the resolution of the VLA (for the NVSS in D/DnC arrays) and ROSAT being 45$\arcsec$ (at 1.4\,GHz) and $\sim$25$\arcsec$ (at 1\,keV) respectively, neither observation was capable of adequately resolving any jets in NGC\,6217. More recently, \citet[][hereafter F17]{Falocco2017}, analysed \textit{XMM-Newton}\footnote{Based on observations obtained with \textit{XMM-Newton}, an ESA science mission with instruments and contributions directly funded by ESA Member States and NASA} images and spectra of NGC\,6217, obtained in 2001, 2002 and 2007, and reported on the alignment of three X-ray `knots' and hence a `likely' X-ray jet emanating from the core in a south-west direction extending over 3$\arcmin$. They supported the original PA01 assertion from the poorer angular resolution X-ray data. However, both PA01 and F17 used the same NVSS images of NGC\,6217. Higher resolution radio data from sub-arcsecond resolution radio interferometers are required to analyse the radio emission in the putative X-ray jet in this galaxy. Hence, interferometers like e-MERLIN and the NSF Karl G. Jansky Very Large Array\footnote{The National Radio Astronomy Observatory is a facility of the National Science Foundation operated under cooperative agreement by Associated Universities, Inc.} (VLA) are ideal instruments for the study of the radio emission and the nature of the aligned X-ray knots in NGC\,6217.

In the literature, NGC\,6217 has not been well observed at radio frequencies other than in surveys with poor angular resolution. The 1465\,MHz radio image of NGC\,6217 at a resolution of $\sim$15$\arcsec$ from \citet{Hummel85} shows an elongation along a P.A. of 160$^{\circ}$, emanating from a core. The follow up survey at lower resolution, $\sim$0.9$\arcmin$, by \citet{Condon87} showed only a core with a weak elongation in the south west (P.A. $\sim$225$^{\circ}$), and no symmetrical counter-jet or outflow. However, \citet{Vila1990} showed a mostly unresolved source at 1.49\,GHz with the VLA in A array, but with a significant extension to the south-east and north-west along a P.A. of $\sim$135$^{\circ}$ and clearly larger than the beam size at 20\,cm of 
$1\farcs5 \times 1\farcs06$. We therefore used the e-MERLIN interferometer as part of the \textbf{L}egacy \textbf{e}-\textbf{M}ERLIN \textbf{M}ulti-band \textbf{I}maging of \textbf{N}earby \textbf{G}alaxies \textbf{s}urvey (LeMMINGs; \citealt{BeswickLemmings}) `deep' sample to take deeper, high-resolution data to directly image the nuclear region at 150\,mas scales at 1.51\,GHz. We also obtained new VLA-A array images to improve the definition of the arcsecond-scale structure.

The distance to NGC\,6217 is listed between $\sim$19 and 35\,Mpc \citep{TullyFisher,Hummel80,Condon90,TutuiSofue,PietschArp,LEDAPaturel2003}. 
We adopt the same distance to NGC\,6217 as F17 of z=0.0045 \cite[20.1\,Mpc with an angular scale of 97\,pc per arcsecond, see also][]{CabreraLavers2004}.
It is of morphological type (R'L)SB(rs)b \citep{RevisedShapley-Ames,deVaucouleursRedshift,Buta2015} i.e. a barred spiral galaxy with an inner and outer (pseudo-) ring structure and with a star formation rate of $\sim$2.9\,M$_{\odot}$ per year \citep{James2004,Knapen2009}. Different authors classify the nuclear region in NGC\,6217 variably as a Seyfert 2 \citep{CabreraLavers2004}, H~II Region \citep{Ho97a, Veron06} or a LINER \citep{Nicholson1997, Falocco2017}. The exact nuclear classification is important for understanding the ionisation mechanisms in the nucleus: Seyferts are ionised by an AGN, whereas H~II regions are ionised through SF processes. On the other hand, LINERs are thought to be ionised by post-AGB stars, LLAGN or shocks, \citep{Heckman80b,Halpern1983,FerlandNetzer1983,TerlevichMelnick1985,CapettiBaldi2011,Singh2013}. The mass of the black hole in NGC\,6217 is estimated from the M-$\sigma$ relation \citep{Magorrian98,FerrareseMerritt00} by using the stellar velocity dispersion of NGC\,6217 from \cite{Ho97b} and using the M-$\sigma$ relation of \cite{WooUrry2002} to obtain a black hole mass of 2.0 $\times$ 10$^6$M$_{\odot}$. 

The purpose of this paper is two-fold: to investigate the nature of the nuclear emission in NGC\,6217 and to check if radio emission in the galaxy spatially is coincident with the X-ray jet suggested by PA01 and F17. This paper is structured as follows: we describe the observations and data reduction in Section~2, in Section 3~we show the resulting radio images and compare these to other ancillary wavelength data and we discuss our Spectral Energy Distributions (SEDs) and X-ray fitting techniques. In Section~4, we discuss the results and, finally, in Section~5 we summarise and present our conclusions. We adopt the following cosmology for computing luminosity distances: H$_0$ = 67.3 km s$^{-1}$ Mpc$^{-1}$, $\Omega _{\Lambda}$ = 0.685 and $\Omega _{M}$ = 0.315 \citep{Planck}.

\begin{figure}
\begin{centering}
\subfloat[]{\includegraphics[width=\columnwidth]{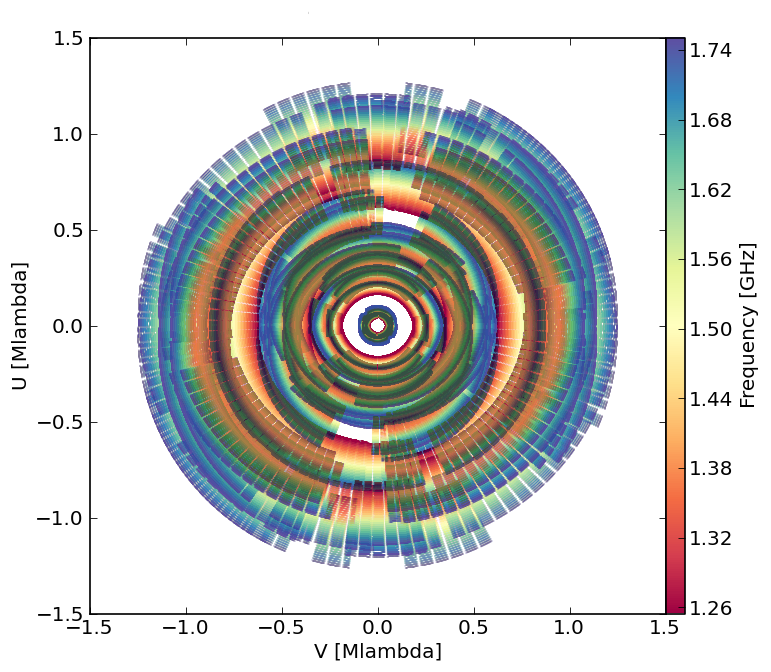}}\\
\hspace*{\fill}
\subfloat[]{\includegraphics[width=\columnwidth]{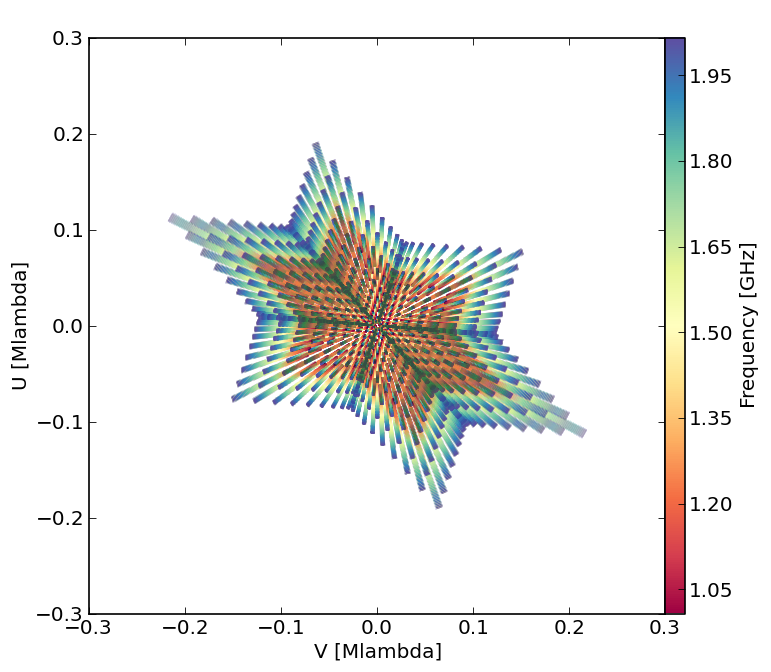}}\\
\hspace*{\fill} 
\caption{Full observed \textit{uv}-plane of NGC\,6217 at 1.51 GHz, using a) the new LeMMINGs (e-MERLIN) deep data with all 7 antennas included and b) the new VLA A-array data with all antennas.}
\label{fig:6217UVPLT}
\end{centering}
\end{figure}

\section{Observations and Data Reduction}
\label{sec:ObsDataReduction}

Here, we discuss the main data sources relevant to this paper: the recently acquired full-resolution 1.51\,GHz e-MERLIN, and VLA data, \textit{XMM-Newton} and other ancillary wavelength data.

\subsection{Radio Data Reduction and Imaging}

\subsubsection{e-MERLIN}
\label{eMERLINData}
The e-MERLIN L-band (1.51\,GHz) observations of NGC\,6217 were obtained as part of the LeMMINGs survey \citep{BeswickLemmings,BaldiLeMMINGs}. LeMMINGs is a statistically-complete radio survey of all the galaxies in the Palomar spectroscopic bright galaxy sample \citep{Filippenko85,Ho95,Ho97a,Ho97b,Ho97e,Ho97c,Ho97d,Ho03,Ho09} of $\delta$ $\ge 20^{\circ}$. The observations were made as part of the `deep' part of the LeMMINGs sample, which constitutes 6 objects of scientific interest, observed at a greater depth than the `shallow' sample and with a fully sampled \textit{uv}-plane: M\,82 \citep{MuxlowM82}, IC\,10 \citep{Westcott2017}, NGC~4151 \citep{Williams4151}, NGC\,5322 \citep{Dullo5322}, M\,51b \citep{RampadarathM51}, and NGC\,6217 (this paper). The data were obtained on 2016 April 7 with all 7 antennas, including the Lovell Telescope, participating in the observing run and 21 hours on-source time. The calibrator J1723+7653 (J172359.4+765312) was used for phase referencing and OQ208 and 3C286 were used as the band pass and flux calibrators respectively. The target was observed for $\sim$12.5 minutes, alternating with a $\sim$2 minute observation of the phase calibrator. The flux and band pass calibrators were observed near the beginning of the run. The \textit{uv}-plane coverage of e-MERLIN and VLA data made with the \textsc{CASA} e-MERLIN pipeline\footnote{https://github.com/e-merlin/CASA$\_$e$-$MERLIN$\_$pipeline/wiki} is shown in Fig~\ref{fig:6217UVPLT} (see the next section for the VLA data overview).

\begin{table}
	\centering
	\caption{Radio datasets of NGC\,6217 at different frequencies with different arrays used in this paper. The e-MERLIN and VLA-A Array datasets observed in 2016 were made as part of the LeMMINGs `deep' sample.}
	\label{tab:Radiodata}
	\begin{tabular}{lcccccc} 
		\hline\hline
		Data & e-MERLIN & VLA \\
		\hline
		Obs. Date (yyyy-mm-dd) & 2016-04-07 & 2016-12-16 \\
		Central Frequency (GHz) & 1.51 & 1.51 \\
		Bandwidth (GHz) & 0.512 & 1.024 \\
		Time on Source & 21 hours & 4.3 mins \\
		Beam size ($\arcsec \times \arcsec$)&  0.15 $\times$ 0.15 & 1.33 $\times$ 0.84\\
		Noise level ($\mu$Jy/beam) & 7 & 53\\
		\hline
	\end{tabular}
\end{table}

The data were correlated and averaged at Jodrell Bank first before they were flagged with the SERPent \citep{SERPent} flagging code to excise the worst cases of radio frequency interference (RFI) from the data before calibration could begin. The \textsc{AIPS} \citep{AIPS} tasks \verb'SPFLG' and \verb'IBLED' allowed for manual inspection of the data to further flag low-level RFI. The observation was centred on 1.51\,GHz, using a total bandwidth of 512\,MHz in 8 intermediate frequencies (IFs) of 64\,MHz each. In addition to RFI flagging, channels at the ends of each IF and the band were flagged as they showed no coherent phase, as were the first two IFs on all Lovell LL-polarisation baselines, due to the inclusion of a test filter.

The procedure outlined in the e-MERLIN cookbook \citep{eMERLINcookbook} was used to calibrate the data, fitting fringes with \textsc{AIPS} task \verb'FRING', phase calibrating and then applying gain corrections to the antennas. The band pass solutions were found and subsequently applied to the data, before imaging and self-calibration of the phase calibrator. Erroneous data points were flagged out at each part of the process before the calibration files were applied.

After initial calibration in \textsc{AIPS}, the e-MERLIN data were imaged using \verb'IMAGR' and self-calibrated. On all baselines, IF 1 and 2 had a very poor signal-to-noise ratio due to RFI, hence, large parts of these bands were removed before producing final images. Other bright sources of emission in a radius of a few arc minutes from the core of NGC\,6217 were also imaged to improve the image fidelity and so that their side lobes could be removed from the final image of NGC\,6217. The images were naturally weighted to achieve the best signal-to-noise ratio. The data were re-weighted using \verb'REWAY', thus increasing the weighting given to the Lovell antenna which is larger and more sensitive, and which further improved the signal-to-noise ratio. Further self calibration followed and the final image was created when the noise in the image reached 7$\mu$Jy. While initial images were made in \textsc{AIPS}, further image making was processed in \textsc{CASA} \citep{CASA}.

\begin{table*}
	\centering
	\caption{\textit{XMM-Newton} datasets on NGC\,6217 from the HEASARC archive relevant to this paper. The observation ID, observation date, instruments available in each observation, exposure time and P.I. of each observation. Notes: $^a$ The PN was turned off for this observation, $^b$ The source fell on the edge of the chip for MOS1, $^c$ The source fell on the edge of the chip for PN. We refer to the data observed on 2007 February 15 as 2007a, and those observed on the 2007 February 17 as 2007b throughout the rest of this work.}
	\label{tab:XrayData}
	\begin{tabular}{lcccccccr} 
		\hline\hline
		Obs. ID & Obs. Date & Exposure &   & Available Instruments &   &   \\
		  & dd-mm-yyyy & ks & Core & Knot 1 & Knot 2 & Knot 3 \\
		\hline
		0061940301 & 20-09-2001 & 6.1  & MOS2$^{a,b}$ & MOS1+2$^a$ & MOS1+2$^a$ & MOS1+2$^a$ \\
		0061940901 & 11-04-2002 & 9.7  & MOS1+2/PN & MOS1+2/PN & MOS1+2$^c$ & MOS1+2$^c$ \\
		0400920101 & 15-02-2007 & 40.4  & MOS1+2/PN & MOS1+2$^c$ & MOS1+2/PN & MOS1+2/PN \\
		0400920201 & 17-02-2007 & 38.9  & MOS1+2/PN & MOS1+2$^c$ & MOS1+2/PN & MOS1+2$^c$ \\
		\hline
	\end{tabular}
\end{table*}

\subsubsection{VLA-A Array data}
\label{VLAData}
NGC\,6217 was observed with the VLA in A configuration on 2016 December 16 for 4.3 minutes centred at 1.51 GHz (L band) over a bandwidth of 1.024\,GHz in 16 spectral windows of 64\,MHz, each in turn with 64 channels. Standard VLA flagging and calibration routines were performed using version 4.7 of the VLA pipeline in \textsc{CASA} 4.7\footnote{https://science.nrao.edu/facilities/vla/data-processing/pipeline}. The pipeline products were reviewed and additional RFI excision was made where necessary. Due to excessive RFI contamination, spectral window 8 was flagged and removed before the final image making procedures. The resulting image achieved an rms of 53$\mu$Jy/beam.

The VLA data were imaged in \textsc{CASA} using \verb'clean', first with a large field to help identify the bright sources in the field and then with smaller fields to remove the effects of these contaminating sources of emission from the NGC\,6217 image. The data were naturally weighted to ensure good signal to noise. 

\subsubsection{Data combination}

After the data were calibrated, we exported the e-MERLIN and VLA datasets to \textsc{CASA} and combined them. First, both datasets were re-weighted using the \textsc{CASA} task \verb'statwt' to re-weight each individual visibility by its rms value. Following this step, both datasets were combined and imaged with \verb'tclean' in \textsc{CASA} with clean boxes specified around the core and all radio sources identified from the e-MERLIN and VLA images and a Briggs weighting scheme (\verb'robust = 0') to image both the compact and diffuse structure. We attempted cleaning with different values of the \verb'multiscale' parameter and the robustness in the Briggs parameter and found the default values of \verb'robust = 0' and \verb'multiscale = [0,5,10]' to give the best compromise between diffuse structure and compact, point-like sources. To properly sample the hybrid point spread function (psf) of the combined e-MERLIN and VLA data, we oversampled the beam by a factor of sixteen (Jack Radcliffe, priv comm). This approach allowed us to calculate the correct hybrid psf which we then used as the \verb'bmaj', \verb'bmin' and \verb'bpa' parameters in \verb'tclean'. We could then image with standard Nyquist sampling while correctly fitting the hybrid psf.

\subsection{XMM-Newton Data Reduction}
\label{sec:XrayData}
The \textit{XMM-Newton} X-ray data were originally published in F17 and we refer to that paper for details. We outline the datasets in Table~\ref{tab:XrayData}, as we do not fit the data that falls on the edges of the chips, for the core and all the knots. 
The data reduction was performed using the standard reduction procedures with version 17 of the Science Analysis System (SAS)\footnote{https://www.cosmos.esa.int/web/xmm-newton/what-is-sas} with CCF release 358. To produce an X-ray image, we aligned and stacked the MOS1, MOS2 and PN event files from all datasets using the task \verb'evselect'  to choose the spectral windows and combined all the epochs and instruments with the FTOOLS task \verb'XSELECT' . We extracted the X-ray surface brightness profiles of the knots and show updated knot centroid positions in Table~\ref{tab:SEDFits}. We find that the positions differ from those in F17 by 2-7$\arcsec$, which can possibly be explained by blending of several sources in a crowded field. To have commensurate source extraction regions with F17, we used a source extraction radius of 25$\arcsec$ around the nucleus and removed the background from a region of the same size from a nearby area void of sources and on the same chip. For the individual knots, we employed the same strategy but with a source region radius of 21$\arcsec$. We checked for pile-up with \verb'epatplot' but found no significant contribution to the spectra over any epoch, chip or source.  
We corrected the X-ray data for Galactic absorption in the direction of NGC\,6217 of 3.90 $\times$ 10$^{20}$ cm$^{-2}$ \citep{Kalberla2005}. As other optical data are available at better than the 4$\arcsec$ angular resolution of the optical monitor (OM), we did not reduce the OM data for these \textit{XMM} observations (see Section~\ref{sec:AncillaryData}). Due to low count rates on this galaxy, we did not analyse the RGS data.

\subsection{Ancillary Data}
\label{sec:AncillaryData}

As NGC\,6217 is part of the Palomar sample \citep{Filippenko85,Ho95,Ho97a,Ho97b,Ho97e,Ho97c,Ho97d,Ho03,Ho09}, there exists a large amount of ancillary data in optical and infra-red surveys. With the intention of fitting the Spectral Energy Distributions (SEDs) of all sources coincident with the galaxy nuclear and X-ray jet knot emission, we retrieved data from several missions from infra-red to optical: \textit{Spitzer}, \textit{WISE}, SDSS and \textit{HST}.

\begin{figure*}
\begin{centering}
	\includegraphics[width=\textwidth]{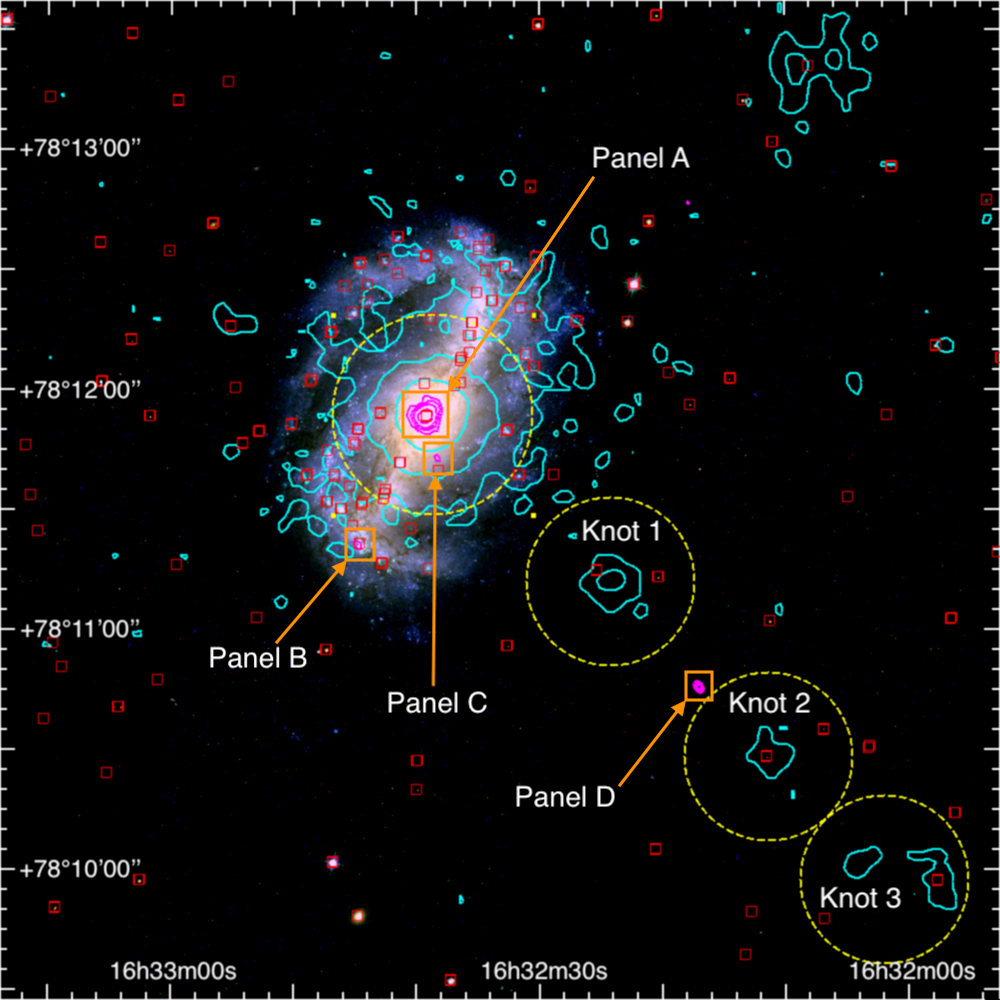}
    \caption{
Multi-wavelength image of NGC\,6217 and the putative 160\,kpc X-ray jet from F17
, including insets of the nuclear region and 3 other sources detected with the combined VLA A-array and e-MERLIN data. 
The background image utilises \textit{HST} F814W (red), F625W (green) and F435W (blue) images to create a three-colour image of NGC\,6217. All three filters are weighted evenly on a linear scale. The cyan colours represent the 7, 12 and 30 count levels of the 0.3-2.0keV \textit{XMM-Newton} image, was smoothed with a Gaussian filter of radius 2$\arcsec$.
The yellow dashed circles correspond to the \textit{XMM-Newton} data spectral extraction regions for the nucleus, and all the X-ray knots from the updated values in Table~\ref{tab:SEDFits}, with radius the same from F17 and in this paper: 25$\arcsec$ for the nucleus and 21$\arcsec$ for each of the knots. The magenta contours show the VLA A-Array data at 0.25, 0.5, 1.0 and 5.0 mJy/beam. The red squares denote 3$\sigma$ detections of sources with the archival 3.6 and 4.5 micron \textit{Spitzer} data. 
The inset panels are described in Figure~\ref{fig:NaturalMapComp} and indicate the ``secondary sources'' described in Section~3.1.2.
Panel A is the nuclear region of NGC\,6217, panel B corresponds to a compact region in the spiral arm of NGC\,6217 south-east of the core, panel C is for a small region of radio emission in the galaxy bulge slightly south of the core and panel D describes a discrete radio source between the X-ray knots 1 and 2, south-west of the core. 
This figure corresponds to a scale of 4$\arcmin$ $\times$ 4$\arcmin$ (25\,kpc $\times$ 25\,kpc).
Note the size of the inset squares overlaid on the HST image are not to scale. In this figure 
north is up and east is left. }
    \label{fig:HSTRadioXMMSpitzer}
\end{centering}
\end{figure*}

There are two observations of NGC\,6217 in the \textit{Spitzer} Heritage Archive\footnote{http://sha.ipac.caltech.edu/applications/Spitzer/SHA/} (AORKEY 30924544, 31025408, PI: K. Sheth). These data were observed at 3.6 and 4.5$\mu$m on 2009 October 27 and 2009 November 27. We downloaded the already reduced data from the S4G Survey \citep{Sheth2010} on the NASA Extragalactic Database (NED) website\footnote{https://ned.ipac.caltech.edu/}. We then used the software \verb'SExtractor'\footnote{https://www.astromatic.net/software/sextractor} to extract all sources in the 3.6 and 4.5$\mu$m images above a 3$\sigma$ threshold. We also downloaded the archival Wide-field Infra-red Survey Explorer (\textit{WISE}) data from the IRSA \footnote{http://irsa.ipac.caltech.edu/frontpage/} in the four bands \citep[3.4, 4.6, 12, and 22 $\mu$m][]{WISEDesc} to help give better constraints on the SED fits for the core and knots at different wavelengths. We downloaded SDSS archival data from the DR12 Science Archive Server \footnote{https://dr12.sdss.org/fields} in all the Johnson $ugriz$ filters. 
We also obtained high-resolution {\it HST} images of NGC\,6217 from the public Hubble Legacy Archive (HLA\footnote{http://hla.stsci.edu}). We downloaded the HST/ACS F814W, F625W and F435W images (PI: Siranni) taken on the 2009 June 13 and 2009 July 8 to create a three-colour image for reference with regards to the other ancillary data. This colour image is shown with the overlaid contours of the ancillary data and radio images inset in Figure.~\ref{fig:HSTRadioXMMSpitzer}. A H$\alpha$-line image (see Figure 5) was obtained with the HST/ACS, through its F658N filter, and reduced as part of the preparatory work for the paper of \cite{Comeron2010}. For continuum subtraction an ACS F814W broad-band image was used, following procedures in \cite{KnapenStedman2004}, \cite{SanchezGallegoKnapen2012} and \cite{DulloMartinezLombilla2016}.

\section{Data Analysis}
\label{DataAnalysis}
Figure~\ref{fig:HSTRadioXMMSpitzer} shows all of the ancillary data overlaid on the three colour \textit{HST} image (see the figure caption for an explanation of the contours and sources in different bands). Three X-ray sources appear to spatially align at a P.A. of $\sim$225$^{\circ}$ from the X-ray source at the centre of NGC\,6217. While knots 1 and 2 show more compact X-ray emission, knot 3 reveals two separate components which makes the identification of multi-band associations more challenging. 

We searched for possible radio (e-MERLIN and VLA) and IR (\textit{Spitzer}) identification of the knots within the X-ray extraction regions (yellow dashed circles in Figure~\ref{fig:HSTRadioXMMSpitzer}). The putative X-ray jet shows no evident radio components from the VLA A-Array and e-MERLIN images. However we do detect a single compact source at 16$^{\textrm{h}}$32$^{\textrm{m}}$16.99$^{\textrm{s}}$, +78$^{\circ}$10${\arcmin}$45.82${\arcsec}$, between X-ray knots 1 and 2, specifically $\sim$34$\arcsec$ from knot 1 and $\sim$25$\arcsec$ from knot 2. Considering 3$\sigma$-significance \textit{Spitzer} 3.6 and/or a 4.5$\mu$m sources, we detected one or two IR associations for each knot within the 21$\arcsec$ X-ray extraction regions. For knots 1 and 2, we only consider the nearest IR source ($<$5$\arcsec$) to the centroid of the X-ray knot as possible identification. For knot 3, one \textit{Spitzer} source lies on one of two distinct components of the X-ray knot (with a separation larger than \textit{XMM-Newton} resolution, see details in Sec.~\ref{sec:SED}). This IR identification is much weaker than the other two X-ray knots. 

We also show other sources of radio emission in the inset boxes in Figure~\ref{fig:HSTRadioXMMSpitzer}, hereafter called secondary sources, with the combined e-MERLIN and VLA data. Inset panel A shows the nuclear region, inset panel B shows a small source south-east of the nucleus on the spiral arm of NGC\,6217, inset panel C shows a region of radio emission south of the core and in the bulge of the galaxy, and panel D shows the source along the same P.A. as the putative X-ray jet mentioned previously, hereafter known as component R for SED fitting purposes (see Sec.~\ref{sec:SED}). 

\subsection{The new e-MERLIN and VLA Data}
\label{sec:RadioData}

\begin{figure*}
\centering
{\includegraphics[width=0.95\columnwidth]{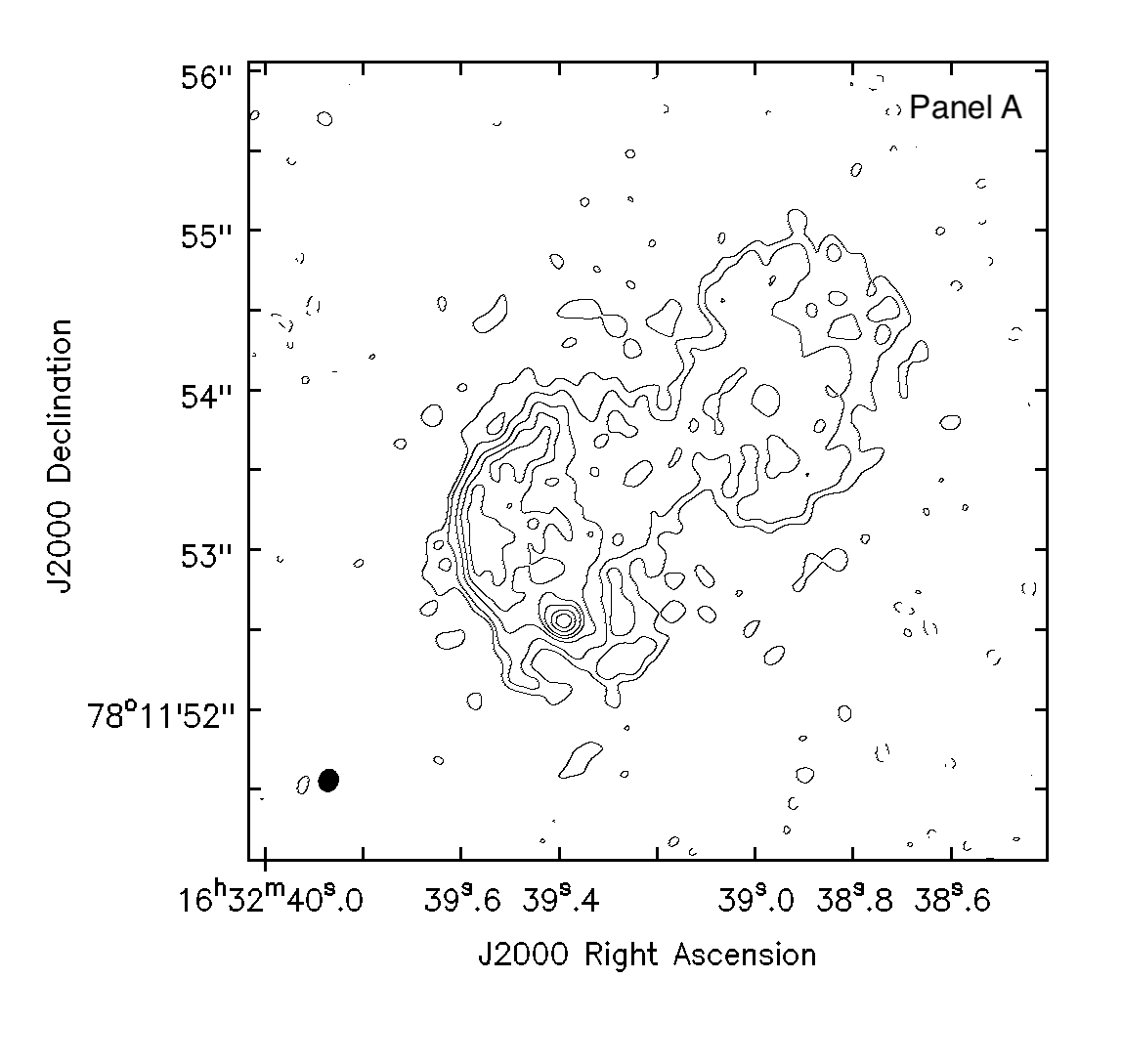}}
{\includegraphics[width=0.95\columnwidth]{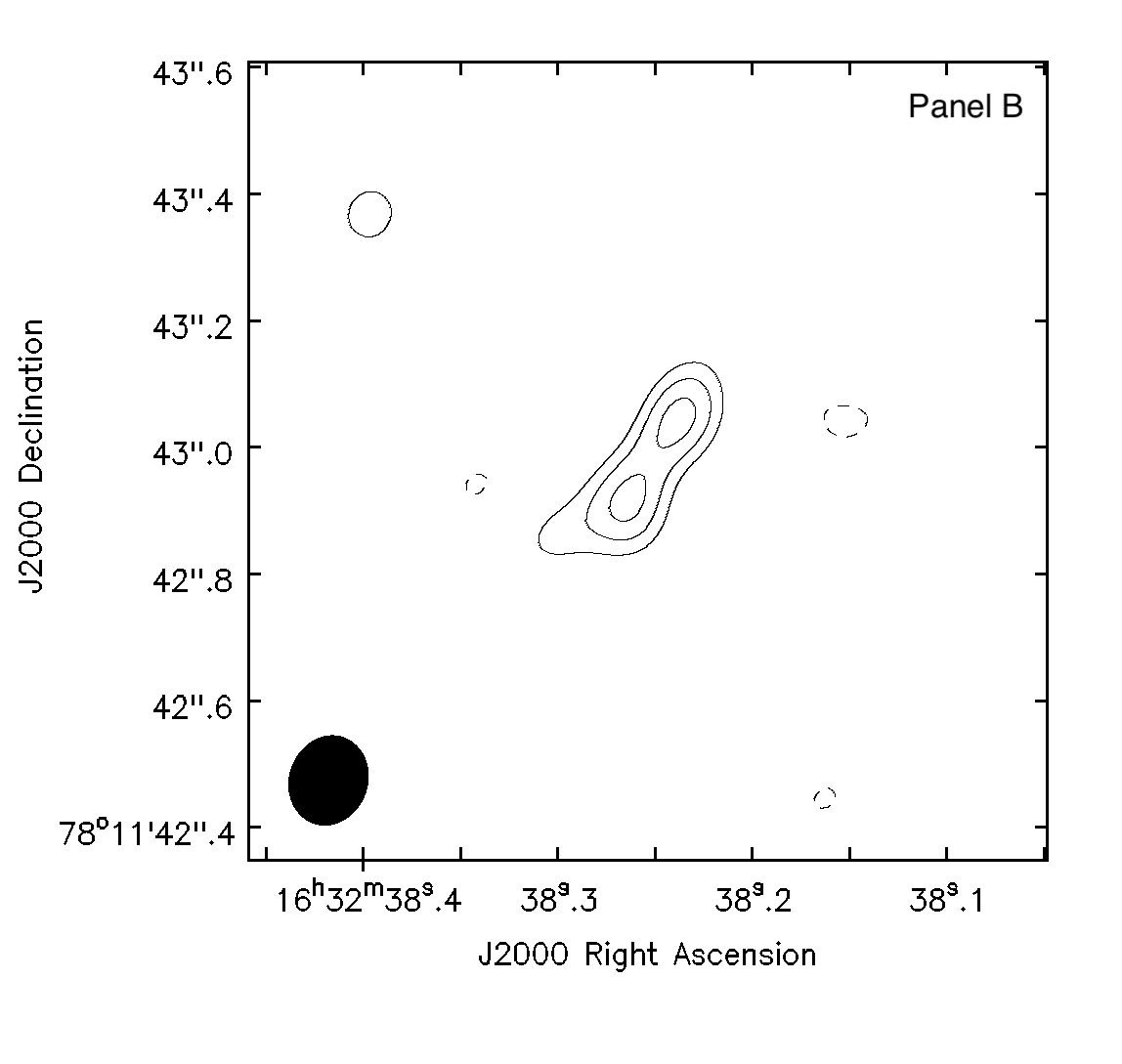}}\\
\centering
{\includegraphics[width=0.95\columnwidth]{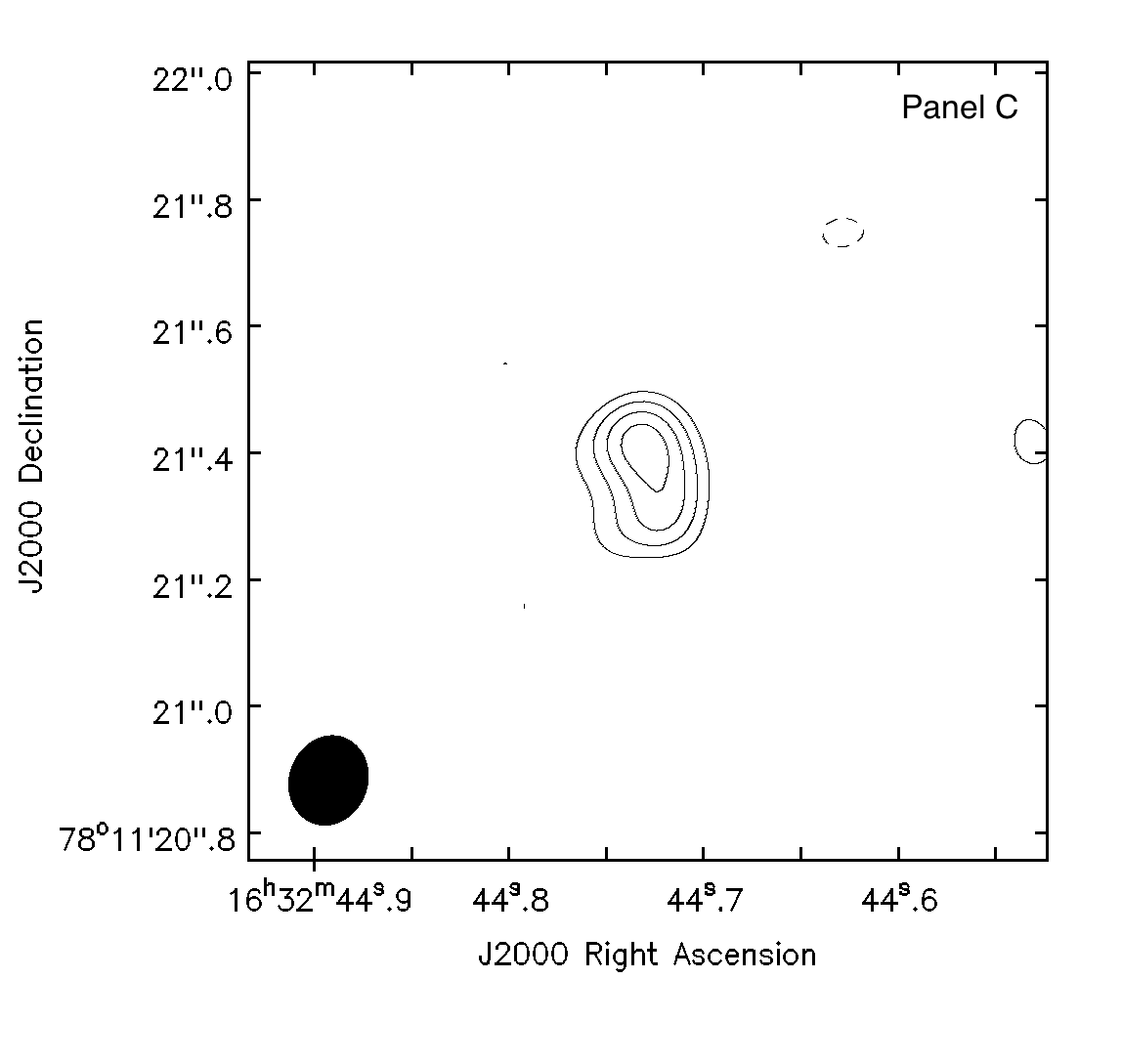}}
{\includegraphics[width=0.95\columnwidth]{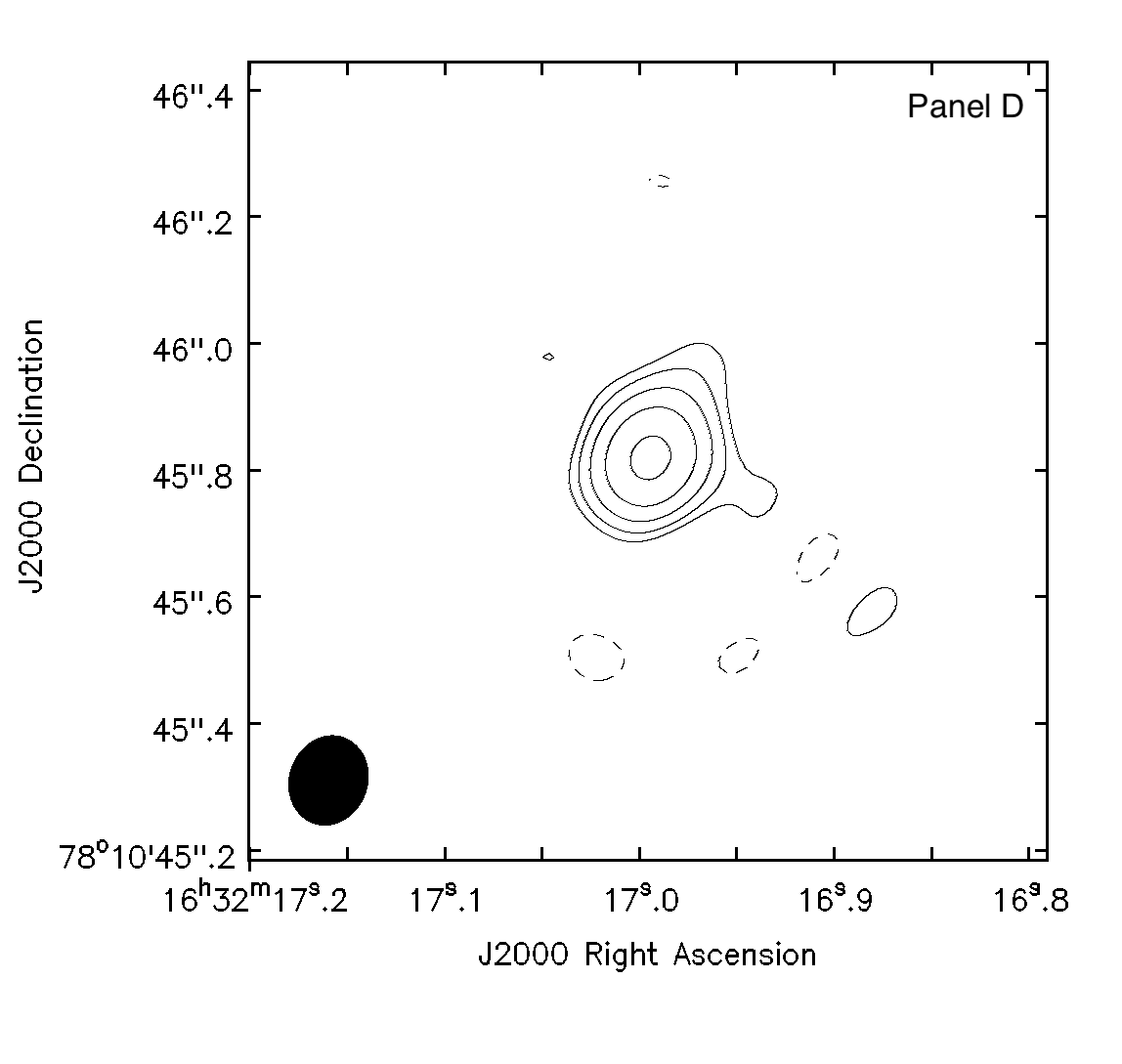}}\\
\caption{Combined e-MERLIN/VLA A-Array, Briggs-weighted contour images of the sources detected with the VLA. The letters in the top-right of each panel correspond to the panels in Fig.~\ref{fig:HSTRadioXMMSpitzer}. The top left is the central 5 $\arcsec$ $\times$ 5 $\arcsec$, corresponding to 485 $\times$ 485 pc, nuclear region in NGC\,6217, as previously shown in panel A of Fig.~\ref{fig:HSTRadioXMMSpitzer}. The combined beam ($0\farcs14 \times 0\farcs12$)  is shown in the bottom left hand corner as a filled black circle. The contour scale is -20, 30, 50, 100, 150, 200, 300, 400, 500 $\times$ 10$^{-6}$Jy beam$^{-1}$. The top right image is the radio source south-east of the nuclear core of NGC\,6217, showing a region $1\farcs2 \times 1\farcs2$, corresponding to 116 $\times$ 116 pc, shown previously in panel B of Fig.~\ref{fig:HSTRadioXMMSpitzer}. The beam is $0\farcs14 \times 0\farcs12$ and the contour scale is -20, 30, 40, 50 $\times$ 10$^{-6}$Jy beam$^{-1}$. The bottom left image is the radio source directly south of the nuclear core of NGC\,6217, shown in panel C of Fig.~\ref{fig:HSTRadioXMMSpitzer}. The beam ($0\farcs14 \times 0\farcs12$) and the contour scale is -20, 30, 40, 50, 60 $\times$ 10$^{-6}$Jy beam$^{-1}$. The bottom right image is the radio source south-east of the nuclear core of NGC\,6217 and in between X-ray knots 1 and 2, shown in panel D of Fig.~\ref{fig:HSTRadioXMMSpitzer}. The contour scale is -50, 50, 100, 200, 400, 800 $\times$ 10$^{-6}$Jy beam$^{-1}$ and the beam size is $0\farcs14 \times 0\farcs12$. In all contour plots, north is up and east is to the left.}
\label{fig:NaturalMapComp}

\end{figure*}

\subsubsection{Radio Morphology of the nucleus}

The upper left panel of Fig.~\ref{fig:NaturalMapComp} shows the 10 $\times$ 10$\arcsec ^2$ nuclear region of NGC\,6217 for the combined 1.51\,GHz e-MERLIN and VLA A-Array data, hereafter referred to as the `combined' data, with `Briggs' weighting to show both the diffuse emission provided by the VLA and the compact core regions that e-MERLIN is sensitive to. The combined image shows a two-sided radio structure (P.A.$\sim$ 135$^{\circ}$) akin to a `peanut', with no apparent core at the centre of this structure. Resolving structures such as this is only possible with the e-MERLIN interferometer with its unique combination of sub-arcsec and sub-mJy imaging.

The south-eastern lobe-like structure in the combined image is clearly brighter than the north-western one, with an integrated flux density of 18.74$\pm$0.24\,mJy compared to 11.60$\pm$0.22\,mJy. There is a hot-spot at 16$^{\textrm{h}}$32$^{\textrm{m}}$39.39$^{\textrm{s}}$, +78$^{\circ}$11${\arcmin}$52.6${\arcsec}$. North of this bright point of emission is a bending arc of radio emission extending $\sim$1.5$\arcsec$ north and slightly west. The hot spot in the southern lobe is clearly the brightest source of radio emission in the nuclear region, but it is offset from the nucleus and of similar size to the beam ($0\farcs20 \times 0\farcs14$\,mas, deconvolved from the beam). It has an integrated flux density of 1.16$\pm$0.12\,mJy and peak flux density of 0.451$\pm$0.034mJy/beam. Therefore, it makes up only $\sim$5 per cent of the total flux density of this region when compared with the VLA. The north-western region of emission appears clumpy and not well defined by any one structure. 

\subsubsection{Radio Morphologies of the secondary sources}
The combined e-MERLIN/VLA images show 3 radio sources in the environs of NGC\,6217, which we discuss here. 
First, we see radio emission in the VLA-only and combined images at 16$^{\textrm{h}}$32$^{\textrm{m}}$38.30$^{\textrm{s}}$, +78$^{\circ}$11${\arcmin}$42.9${\arcsec}$ in the spiral arms, seen in inset panel B of Fig.~\ref{fig:HSTRadioXMMSpitzer} and the top right panel in Fig.~\ref{fig:NaturalMapComp}. It has a slightly resolved two-lobed structure. 
The VLA-only data shows a mostly unresolved source, with flux densities of 0.317$\pm$0.041\,mJy/beam (peak) and 0.597$\pm$0.114\,mJy (integrated). This source has a 3$\sigma$ \textit{Spitzer} association.

The second source (inset panel C in Fig.~\ref{fig:HSTRadioXMMSpitzer}  and the bottom left panel in Fig.~\ref{fig:NaturalMapComp}) we observe is in projection within the main galactic bulge, only $\sim$0.5$\arcmin$ south of the nucleus (16$^{\textrm{h}}$32$^{\textrm{m}}$44.73$^{\textrm{s}}$, +78$^{\circ}$11${\arcmin}$21.4${\arcsec}$). It appears as a compact source in the VLA-only and combined images, with a possible small extension east, although given the beam-size and resolution, it is consistent with being an unresolved source. It also has low VLA flux densities of 0.556$\pm$0.044\,mJy/beam (peak) and 0.501$\pm$0.074\,mJy (integrated).

Interestingly, we find a prominent unresolved source of radio emission at 16$^{\textrm{h}}$32$^{\textrm{m}}$16.99$^{\textrm{s}}$, +78$^{\circ}$10${\arcmin}$45.82${\arcsec}$, along the same P.A. as the putative X-ray jet between knots 1 and 2 of the X-ray emission (see panel D of Fig.~\ref{fig:HSTRadioXMMSpitzer} and bottom right panel in Fig.~\ref{fig:NaturalMapComp}). The VLA integrated flux density of this source is 1.34$\pm$0.16\,mJy, which compares to 0.974$\pm$0.036\,mJy with e-MERLIN. This source was not detected above the 3$\sigma$ threshold by the \textsc{SExtractor} procedure, but a weak \textit{Spitzer} source is clearly visible in the image (see Section~\ref{sec:SED}).

\subsubsection{Radio Properties and Spectra}

\begin{table}
	\centering
	\cprotect\caption{Radio flux densities of the core region (panel A in Fig.~\ref{fig:HSTRadioXMMSpitzer}) and all three secondary sources in the inset panels of Fig.~\ref{fig:HSTRadioXMMSpitzer}. All values were obtained from Briggs weighted data, and were made using the \textsc{CASA} 2D fitter where possible, whereas the e-MERLIN integrated flux densities for the core and the individual lobes were obtained using \verb'TVSTAT' in \textsc{AIPS} due to their being extended in nature. The peak flux density of the core region in the e-MERLIN data was not estimated as it is a resolved source and no radio core is observed between the two e-MERLIN radio lobes. The integrated flux for the entire nuclear region was extrapolated from all of the emission from both the lobes, as the lobes are resolved at e-MERLIN and e-MERLIN/VLA baselines. The e-MERLIN data did not detect the sources in panel B and C alone, hence they have no flux values given, but must be less than 3 times the noise in the image (0.02mJy). The columns are presented as follows: (1) is the radio source, (2) is the radio array used, (3) is the integrated flux density from the whole source in the given array, (4) is the peak flux density from the source.}
	\label{tab:RadioComparison}
	\begin{tabular}{cccccr} 
		\hline\hline
		Radio & Radio & Int. Flux & Peak Flux   \\
		Source & Array & (mJy) & (mJy/beam)  \\
		 (1) & (2) & (3) & (4)\\
		\hline
		Total & NVSS & 80.3$\pm$2.4 & 57.8$\pm$1.1\\
		\hline
		Northern & VLA & - & -\\
		Lobe & e-MERLIN & 11.60$\pm$0.22 & 0.146$\pm$0.003\\		
		\hline
		Southern & VLA & - & -\\
		Lobe & e-MERLIN & 18.74$\pm$0.24 & 0.205$\pm$0.003\\
		\hline
		Southern & VLA & - & -\\
		Hot-spot & e-MERLIN & 1.16$\pm$0.12 & 0.451$\pm$0.034\\			
		\hline
		Nuclear & VLA & 40.4$\pm$3.2 & 7.28$\pm$0.50\\
		Structure & e-MERLIN & 29.84$\pm$0.16 & - \\
		\hline
		\hline
		Source in & VLA & 0.597$\pm$0.114 & 0.317$\pm$0.041\\
		Panel B & e-MERLIN & - & - \\
		 \hline
		Source in & VLA & 0.501$\pm$0.074 & 0.556$\pm$0.044\\
		Panel C & e-MERLIN & - & - \\
		 \hline
		Source in & VLA & 1.34$\pm$0.16 & 1.297$\pm$0.088\\
		Panel D & e-MERLIN & 0.974$\pm$0.036 & 0.955$\pm$0.020\\
		\hline
	\end{tabular}
\end{table}

The nuclear region appears resolved in the combined data, hence it is possible that we may have lost some flux due to resolving out the diffuse emission that e-MERLIN is not sensitive to. Table~\ref{tab:RadioComparison} shows the flux densities for the core and all of the other radio sources found in the e-MERLIN-only and VLA-only data. We additionally compare the nuclear region to the NVSS values of NGC\,6217 to fully consider any diffuse emission in the galaxy. From the NVSS and VLA data, we conclude that we lost $\sim$50 per cent of the total emission moving to the longer baseline array. We find a similar drop when comparing the VLA to the e-MERLIN data. Hence, there is diffuse emission that e-MERLIN alone cannot account for in this nuclear region. 

As for the three discrete sources of radio emission seen in the VLA images other than the nuclear region, the two sources in panel B and C in Fig.~\ref{fig:HSTRadioXMMSpitzer} seem to significantly resolve out flux when the e-MERLIN-only (or combined) data is compared to the VLA-only data. For these two sources, e-MERLIN only captures $\sim$40 per cent of the total flux for these regions seen in the VLA-A array data. Hence, the diffuse elongations seen by the combined e-MERLIN/VLA data may in fact be real for these sources. For the third source in panel D in Fig.~\ref{fig:HSTRadioXMMSpitzer}, e-MERLIN data reproduces over 80 per cent of the total flux density of this source when compared to the VLA data, indicating it is likely a very compact source.

We checked for any variability in the core between the \citet{Vila1990} observations and our new VLA-A array data. The flux densities of our new VLA A-array data and of \citet{Vila1990} agree within errors, indicating that there has likely been no significant change over the period of nearly 30 years. 

Due to the low signal for the nuclear region, it was not possible to make an in-band spectral index image from the 1.51\,GHz e-MERLIN observations. However, using the VLA data at 1.51\,GHz, we were able to get a crude in-band spectral index using \textsc{CASA} and by setting \verb'nterms = 2'. We estimate a flat spectral index of $-$0.2 (S$_{\nu}$ $\sim$ $\nu ^\alpha$), although the uncertainties in this value are likely of order $\pm$0.5, as NGC\,6217 is not bright enough in the VLA data to properly constrain $\alpha$. We note that \cite{Vila1990} estimate a compact nuclear spectral index of $-$1.0 based on archival VLA data at 20\,cm and 6\,cm, albeit with a larger beam size. 

\subsection{Spectral Energy Distributions}
\label{sec:SED}

\begin{table*}
	\centering
	\caption{The multi-band magnitudes and fluxes of the sources found coincident with the three X-ray knots and the radio knot (Panel D of Fig.~\ref{fig:HSTRadioXMMSpitzer}). The positions of the X-ray knots and \textit{Spitzer} sources are given, as well as the difference in position, $\Delta$, which was calculated from the difference of the \textit{Spitzer} source position and our X-ray knot centroids. The SDSS, \textit{HST} and \textit{WISE} values are given in AB magnitudes and the IRAC data are given in mJy. The photometric redshift and corresponding bolometric luminosities are given for all of the sources, ascertained from the \textit{2SPD} fitting code of \citet{BaldiSEDFitsa} and \citet{BaldiSEDFitsb}. The Photon Index, $\Gamma$, and the corresponding fluxes and luminosities of each of the X-ray knots is also given, found from the X-ray spectral fitting process in Section~\ref{sec:X-rayDataAnalysis}. We show the X-ray luminosities given the distance to NGC\,6217 (see Section~\ref{sec:Introduction}) and the photometric redshifts ascertained from the SED fits. The errors on the X-ray luminosities include only the uncertainties in the fluxes, and not the distance.}
	\label{tab:SEDFits}
	\begin{tabular}{lccccc} 
		\hline\hline
	      X-ray knot & 1 & 2 & 3 & radio\\
	      \hline
	      		 RA & 16$^{\textrm{h}}$32$^{\textrm{m}}$24.21$^{\textrm{s}}$  & 16$^{\textrm{h}}$32$^{\textrm{m}}$24.21$^{\textrm{s}}$ & 16$^{\textrm{h}}$32$^{\textrm{m}}$01.93$^{\textrm{s}}$ & --\\
		 DEC & +78$^{\circ}$11${\arcmin}$53.8${\arcsec}$ & +78$^{\circ}$11${\arcmin}$12.0${\arcsec}$ & +78$^{\circ}$09${\arcmin}$57.5${\arcsec}$& --\\
		 \hline
		 Associated source & A & B & C & R   \\
		 \hline
		 RA & 16$^{\textrm{h}}$32$^{\textrm{m}}$25.32$^{\textrm{s}}$ & 16$^{\textrm{h}}$32$^{\textrm{m}}$11.49$^{\textrm{s}}$ & 16$^{\textrm{h}}$31$^{\textrm{m}}$57.66$^{\textrm{s}}$ & 16$^{\textrm{h}}$32$^{\textrm{m}}$16.99$^{\textrm{s}}$ \\
		 DEC & +78$^{\circ}$11${\arcmin}$15.1${\arcsec}$ & +78$^{\circ}$10${\arcmin}$28.4${\arcsec}$ & +78$^{\circ}$09${\arcmin}$57.0${\arcsec}$& +78$^{\circ}$10${\arcmin}$45.8${\arcsec}$ \\
		 $\Delta$ & 4.62$\arcsec$ & 0.48$\arcsec$ &  13.15$\arcsec$& -- \\
		 \hline
		 SDSS $u$ & <22.15 & 22.27$\pm$0.37 & 21.93$\pm$0.24 & <22.20 \\
		 SDSS $g$ & <22.20 & 20.43$\pm$0.03 & 22.00$\pm$0.10 & <22.30 \\
		 SDSS $r$ & <22.40 & 21.91$\pm$0.14 & 21.65$\pm$0.11 & <22.20 \\ 
		 SDSS $i$ & <21.40 & 21.71$\pm$0.17 & 21.32$\pm$0.11 & <21.20 \\ 
		 SDSS $z$ & <21.45 & 20.55$\pm$0.25 & 21.35$\pm$0.42 & <21.60 \\ 
		 \textit{HST} F425W & $-$ & 21.32$\pm$0.15 & 22.10$\pm$0.13 & $<$ 22.71\\
		 \textit{HST} F625W & $-$ & 21.19$\pm$0.14 & 21.55$\pm$0.15 & $<$ 22.21\\
		 \textit{HST} F814W & $-$ & 21.58$\pm$0.14 & $-$            & $<$22.42  \\         
		 \textit{WISE} W1 & 16.02$\pm$0.30 & 17.42$\pm$0.09 & 17.30$\pm$0.07 & 18.03$\pm$0.18 \\ 
		 \textit{WISE} W2 & 16.01$\pm$0.20 & 16.18$\pm$0.10 & 17.13$\pm$0.22 & 17.73$\pm$0.48 \\ 
		 \textit{WISE} W3 & <12.28 & 12.50$\pm$0.27 & <12.96 & <13.20 \\ 
		 \textit{WISE} W4 & <8.19 & 9.02$\pm$0.29 & <9.40 & <9.65 \\
		 \textit{IRAC} 1 & 55.10$\pm$8.72 & 39.34$\pm$4.91 & 29.35$\pm$5.35 & 19.00$\pm$6.80 \\
		 \textit{IRAC} 2 & 58.45$\pm$7.25 & 51.57$\pm$4.23 & 31.11$\pm$4.40 & 22.56$\pm$7.34 \\
		 \hline
		 \rule{0pt}{12pt}$z_{phot}$ & $0.48^{+0.25}_{-0.12}$ & $0.70^{+0.03}_{-0.02}$ & $0.46^{+0.19}_{-0.06}$ & $0.19^{+0.25}_{-0.12}$ \\
		 \rule{0pt}{12pt}$L_{bol}$, ($\times$10$^{43}$ erg s$^{-1}$) & $7.9^{+13.8}_{-3.9}$ & $8.3^{+0.9}_{-0.7}$& $9.3^{+12}_{-2.7}$ & $0.46^{+2.69}_{-0.40}$\\
		 \hline
		 \rule{0pt}{12pt}Photon Index, $\Gamma$ & $1.59^{+0.27}_{-0.25}$ & $1.38^{+0.45}_{-0.37}$ & $1.54^{+0.54}_{-0.46}$ & --\\
		 \rule{0pt}{12pt}Total Flux 0.3-10\,keV ($\times$10$^{-14}$ erg s$^{-1}$ cm$^{-2}$) & 6.92$^{+2.20}_{-1.67}$ & 4.37$^{+3.58}_{-2.46}$ & 4.27$^{+3.32}_{-2.08}$ & $<$0.3\\
		 \rule{0pt}{12pt}Unabs. Flux 0.3-10\,keV ($\times$10$^{-14}$ erg s$^{-1}$ cm$^{-2}$) & 8.13$^{+2.34}_{-1.96}$ & 4.79$^{+3.73}_{-2.65}$ & 4.90$^{+3.42}_{-2.33}$ & $<$0.3\\
		 \hline
		 \rule{0pt}{12pt}$L_{X-ray, Unabs}$ at NGC\,6217 ($\times$10$^{39}$ erg s$^{-1}$) & 3.93$^{+1.13}_{-0.95}$ & 2.32$^{+1.80}_{-1.29}$ & 2.37$^{+1.65}_{-1.13}$ & $<$0.13\\
		 \rule{0pt}{12pt}$L_{X-ray, Unabs}$ at photo-z ($\times$10$^{43}$ erg s$^{-1}$)   & 7.58$^{+2.18}_{-1.83}$ & 17.33$^{+13.51}_{-9.58}$ & 4.13$^{+2.88}_{-1.96}$ & $<$0.13 \\
		\hline
		\hline
	\end{tabular}
\end{table*}

We carried out multi-band source identification of the X-ray knots and the VLA-detected radio source, using SDSS \emph{ugriz} bands, the three \textit{HST} ACS bands (F435W, F625W, and F814W), \textit{Spitzer} 3.6 and 4.5$\mu$m data and the four \textit{WISE} bands. We performed aperture photometry on an area which includes the FWHM of the PSF for each instrument, properly centred on the position of the detected \textit{Spitzer} sources, isolating the genuine emission of the associations and subtracting the background emission. The extracted magnitudes of the detected sources are consistent with the values reported in the IRSA catalogue within the uncertainties. For the sources not detected, we derived 3$\sigma$ upper limits on the multi-band association. The optical-IR magnitudes for our sources are presented in Table~\ref{tab:SEDFits}. 


We fit the SEDs with the \textit{2SPD} code developed by \citet{BaldiSEDFitsa} and \citet{BaldiSEDFitsb}. This code allows for the inclusion of two different stellar populations, typically one younger and one older (YSP and OSP, respectively) which correspond to different starburst ages, a dust component with a single (or two) temperature black-body emission(s), AGN spectrum with emission lines. The code searches for the best match between the sum of the different models and the photometric points minimising the appropriate $\chi^{2}$ function (see \citealt{BaldiSEDFitsa} for details). The main parameters which \textit{2SPD} returns are the photometric redshift and the bolometric luminosity of the source. To estimate the uncertainties on the photo-z and mass derivations, we measure the 99 \textit{per cent} confidence solutions for these quantities.

The synthetic stellar templates used to model the observed SEDs are from \cite{bruzual11}, with single stellar populations with ages ranging from 1\,Myr to 12.5\,Gyr. We adopt a dust-screen model for the extinction normalised with the free parameter A$_{V}$, and the \citep{calzetti00} extinction law. The emission lines are taken from the AGN atlas covering the ultraviolet to near-infrared spectral range from Seyfert/QSO templates\footnote{http://www.stsci.edu/hst/observatory/crds/cdbs\_agn.html} \citep{francis91}.

We run the \textit{2SPD} fitting code for different sources: X-ray knot 1 associated with \textit{Spitzer} source A, X-ray knot 2 associated with \textit{Spitzer} source B, X-ray knot 3 associated with \textit{Spitzer} source C, and then the VLA radio component with \textit{Spitzer} source R. All of the extracted \textit{HST}, \textit{WISE} and SDSS sources for all the X-ray knots are detected above 5$\sigma$ detection significance. For the radio knot, the associated \textit{Spitzer} source has a 2.7$\sigma$ detection in IRAC1 and 3$\sigma$ detection for IRAC2. The results from the SED-fitting procedure are found in Table~\ref{tab:SEDFits}. We did not perform SED fitting for the radio sources in panels B and C in Fig.~\ref{fig:HSTRadioXMMSpitzer} as the multi-band contamination from the host galaxy NGC\,6217 is significant, preventing an accurate extraction of the genuine emission from those sources.

X-ray Knot 1 has two \textit{Spitzer} components within the X-ray spectral extraction region, however we only perform an SED fit for the source (A) at 16$^{\textrm{h}}$32$^{\textrm{m}}$25.32$^{\textrm{s}}$, +78$^{\circ}$11${\arcmin}$15.1${\arcsec}$, which is consistent with the the X-ray knot centroid within the position errors. It is detected only in IRAC bands 1 and 2 (3.5 and 4.5$\mu$m), and \textit{WISE} 1 and 2 bands (3.4 and 4.6$\mu$m). In addition, we detected two \textit{HST} sources 1.2$\arcsec$ and 2.5$\arcsec$ away from this \textit{Spitzer} source, but they are not detected in the SDSS. As it is not certain whether and which of the two HST sources are genuinely related to the \textit{Spitzer} source, we do not include them in the SED fit. Using the only detections from \textit{Spitzer} and WISE W1/W2 bands, \textit{Spitzer} source A has an SED that is consistent with an obscured 6\,Gyr old stellar population. The photometric redshift derived is $0.48^{+0.25}_{-0.12}$ and its bolometric luminosity is 7.9$\times$ 10$^{43}$ erg s$^{-1}$. The paucity of accurate data-points in the SED of this source makes the physical parameters derived from the fit for this source doubtful. Therefore the uncertainties on the the redshift and bolometric luminosity are likely much larger than that quoted here. 

X-ray Knot 2 also has two \textit{Spitzer} associations falling in the X-ray extraction region but the closest to the X-ray knot matches with the \textit{Spitzer} position (separation $<$0.3$\arcsec$) and is named as \textit{Spitzer} source B (16$^{\textrm{h}}$32$^{\textrm{m}}$11.49$^{\textrm{s}}$, +78$^{\circ}$10${\arcmin}$28.4${\arcsec}$). It is detected in all the 14 available photometric bands. Its SED shows a rising blue part, flat IR spectrum and interestingly the \textit{g} and \textit{z} band associations are significantly brighter than the other adjacent SDSS bands by $\sim$1.5 magnitudes. The only physical interpretation of those broad-band peaks is the contribution from emission lines in the band-widths of \textit{g} and \textit{z} bands. The best-fit model of the SED includes a Seyfert/QSO-like spectrum, which is able to reproduce simultaneously the blue spectral rising and emission lines, Mg~II and a H$\beta$, which contribute to $\sim$70 and $\sim$66 per cent of the total flux in the \textit{g} and \textit{z} band, respectively. The wavelength of the emission lines within the SDSS bands constrains the photo-z at $0.70^{+0.03}_{-0.02}$. The near- and mid-IR data are modelled by a 3-Gyr old stellar population and two-temperature black bodies. The bolometric luminosity is 8.3 $\times$ 10$^{43}$ erg s$^{-1}$.

As mentioned previously, the shape of X-ray Knot 3 is convoluted as it is made up of two distinct X-ray components which makes the centroid difficult to determine. The western part of this knot (16$^{\textrm{h}}$31$^{\textrm{m}}$57.66$^{\textrm{s}}$, +78$^{\circ}$09${\arcmin}$57.0${\arcsec}$) aligns well with \textit{Spitzer} source C, which is detected in all bands, except the last two \textit{WISE} bands. The blue spectral rising has been fitted with a YSO of 0.03 Gyr and the optical and near-IR region with a OSP of 4 Gyr. The contribution from an AGN is not needed to fit the optical-IR data-points. As the blue spectral curvature is the only feature to constrain the photometric redshift, this results in a mediocre photo-z determination, corresponding to $0.46^{+0.19}_{-0.06}$. The bolometric luminosity is 9.3 $\times$ 10$^{43}$ erg s$^{-1}$. The tentative identification of the X-ray knot with the \textit{Spitzer} source C must be taken as a reasonable attempt to infer the multi-band properties of this X-ray knot without giving any confidence in its correctness.

We also performed an SED fit for the radio component in the inset panel D in Figure~\ref{fig:NaturalMapComp}, bottom right panel, detected with the VLA and coincident with \textit{Spitzer} source R. Its radio emission is unresolved and lies along the putative `X-ray jet' in NGC\,6217 between knot 1 and 2. Its association with one of the two X-ray knots seems to be unlikely due to its large relative distance ($>$25$\arcsec$). This source is detected at near-infrared wavelengths (two IRAC and two \textit{WISE} bands). The resulting SED fitting returned an obscured 4 Gyr old stellar population with a photometric redshift of $0.19^{+0.25}_{-0.12}$ and a bolometric luminosity of 4.6 $\times$ 10$^{42}$ erg s$^{-1}$. The small number of photometric data-points reflects on a poor accuracy of its photo-z determination. Therefore the values inferred for this component must be considered with caution and are meant to provide a possible interpretation of this radio component.

\begin{figure*}
\centering
{\includegraphics[width=\columnwidth]{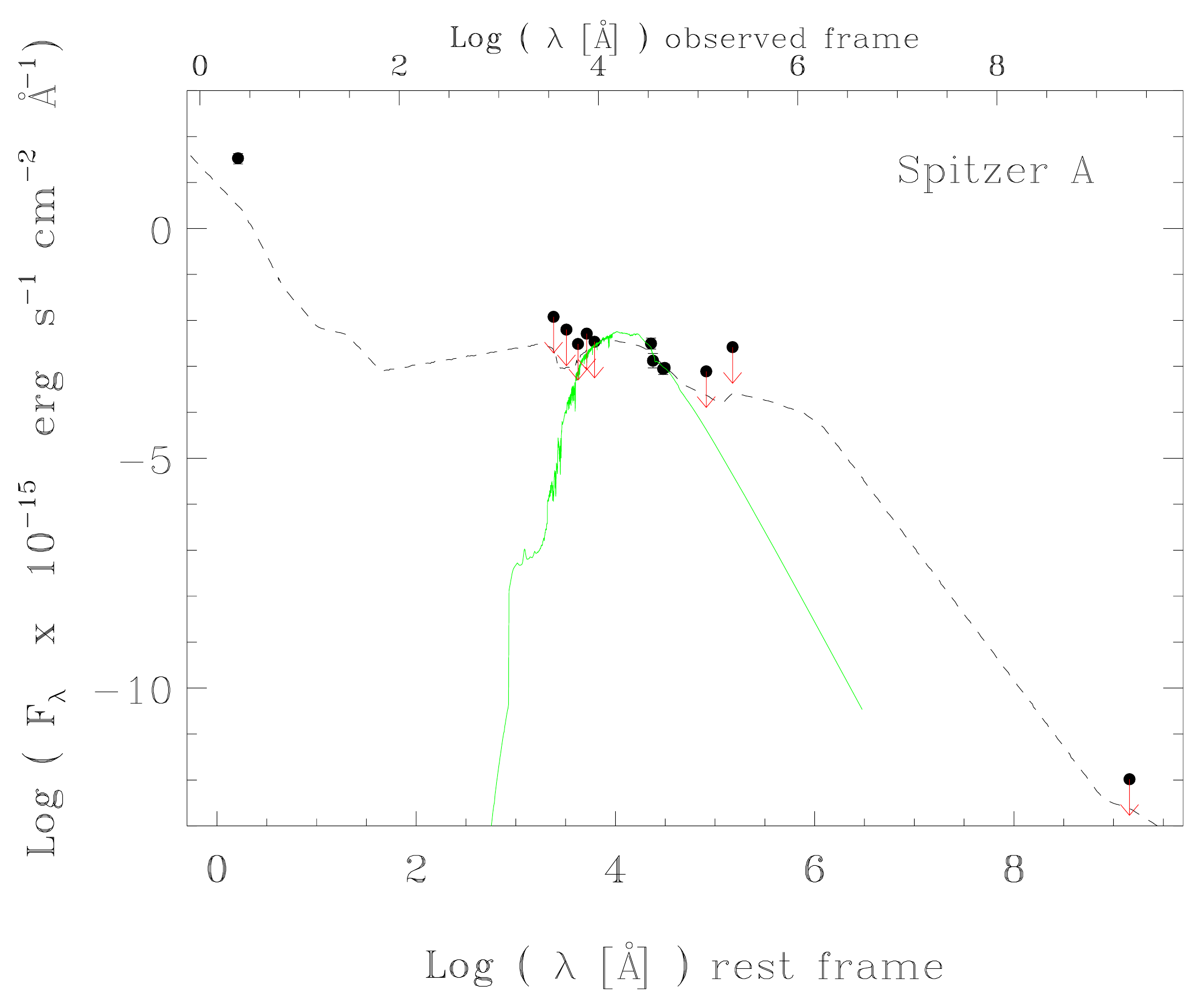}}
{\includegraphics[width=\columnwidth]{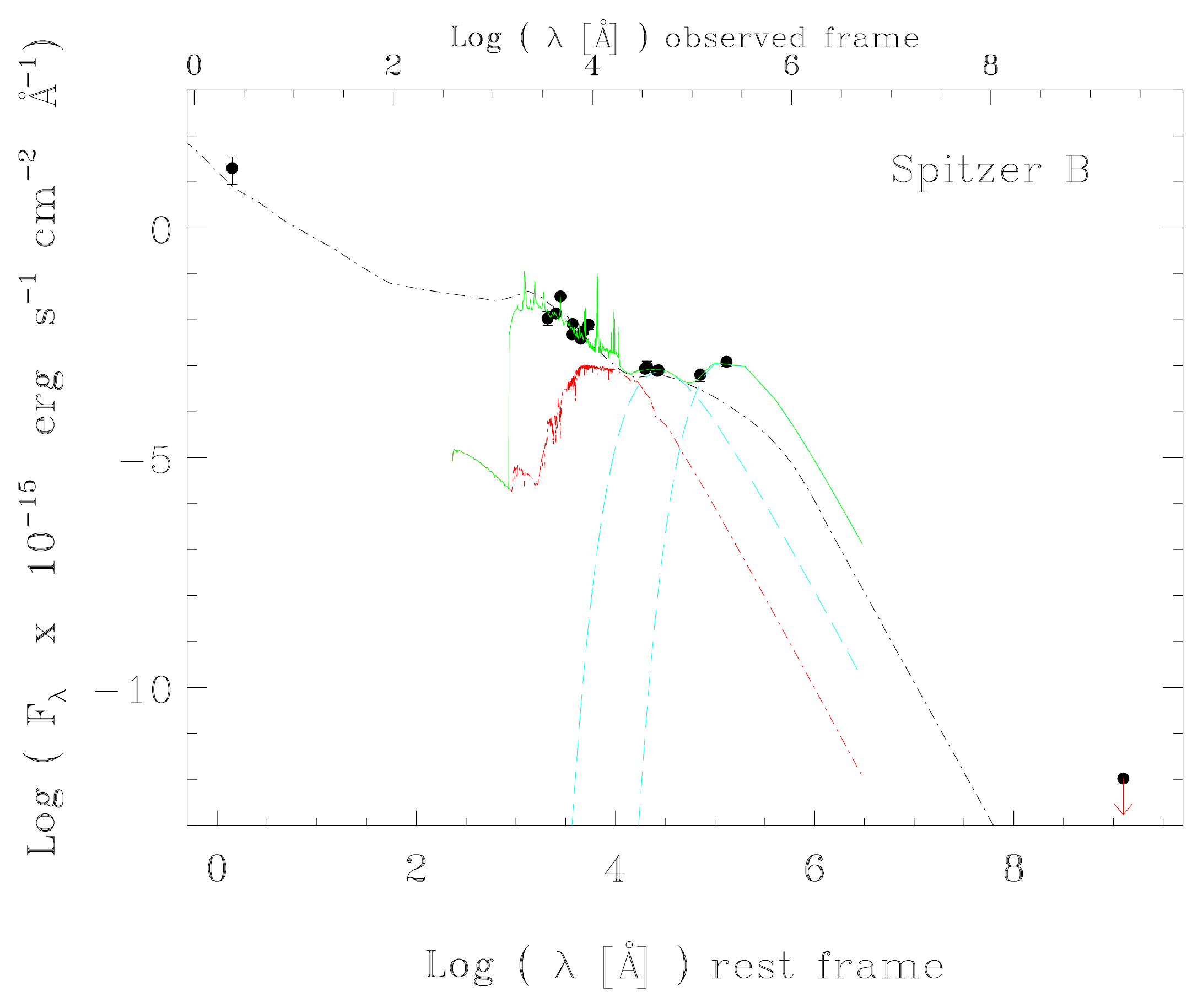}}\\
\centering
{\includegraphics[width=\columnwidth]{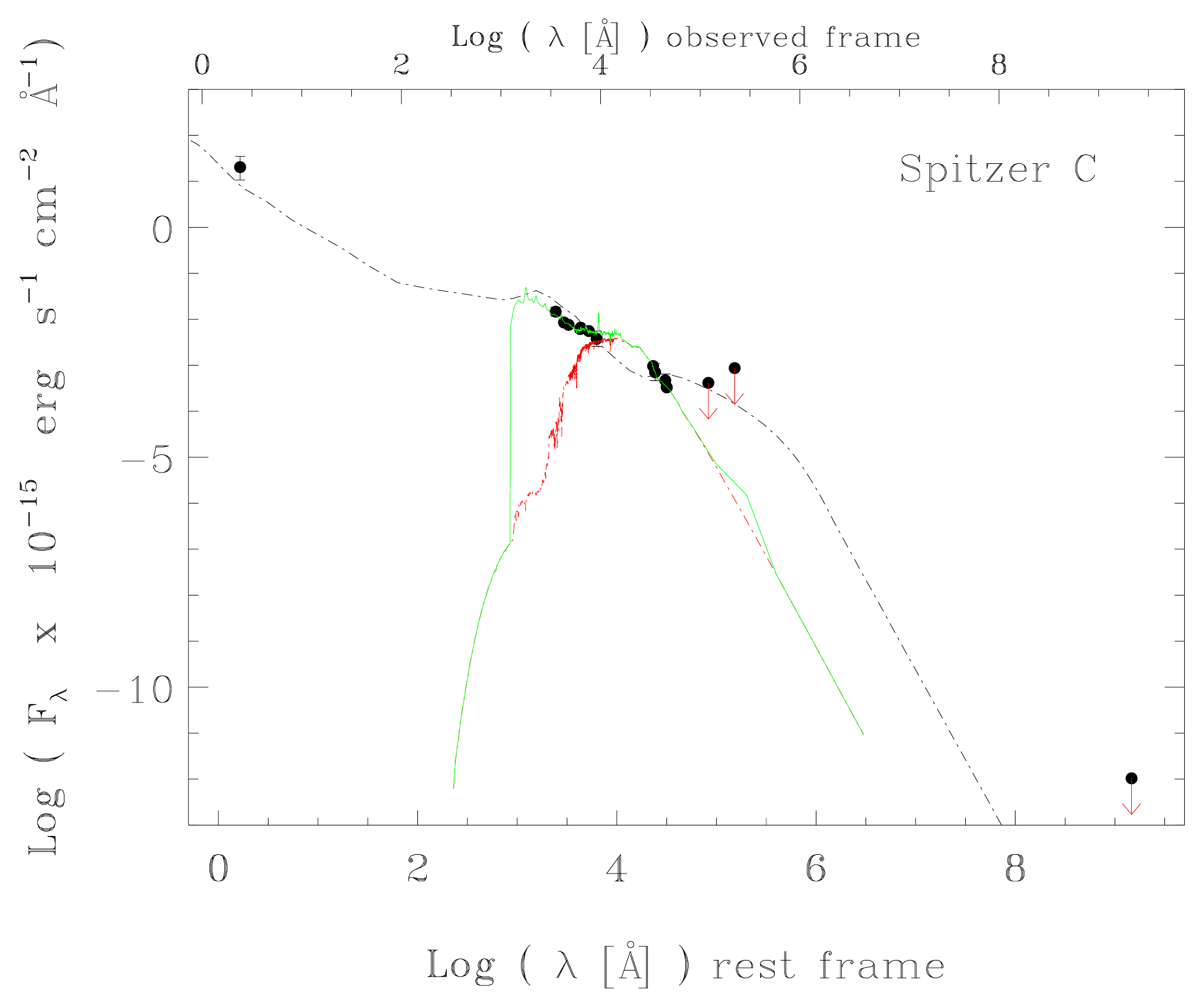}}
{\includegraphics[width=\columnwidth]{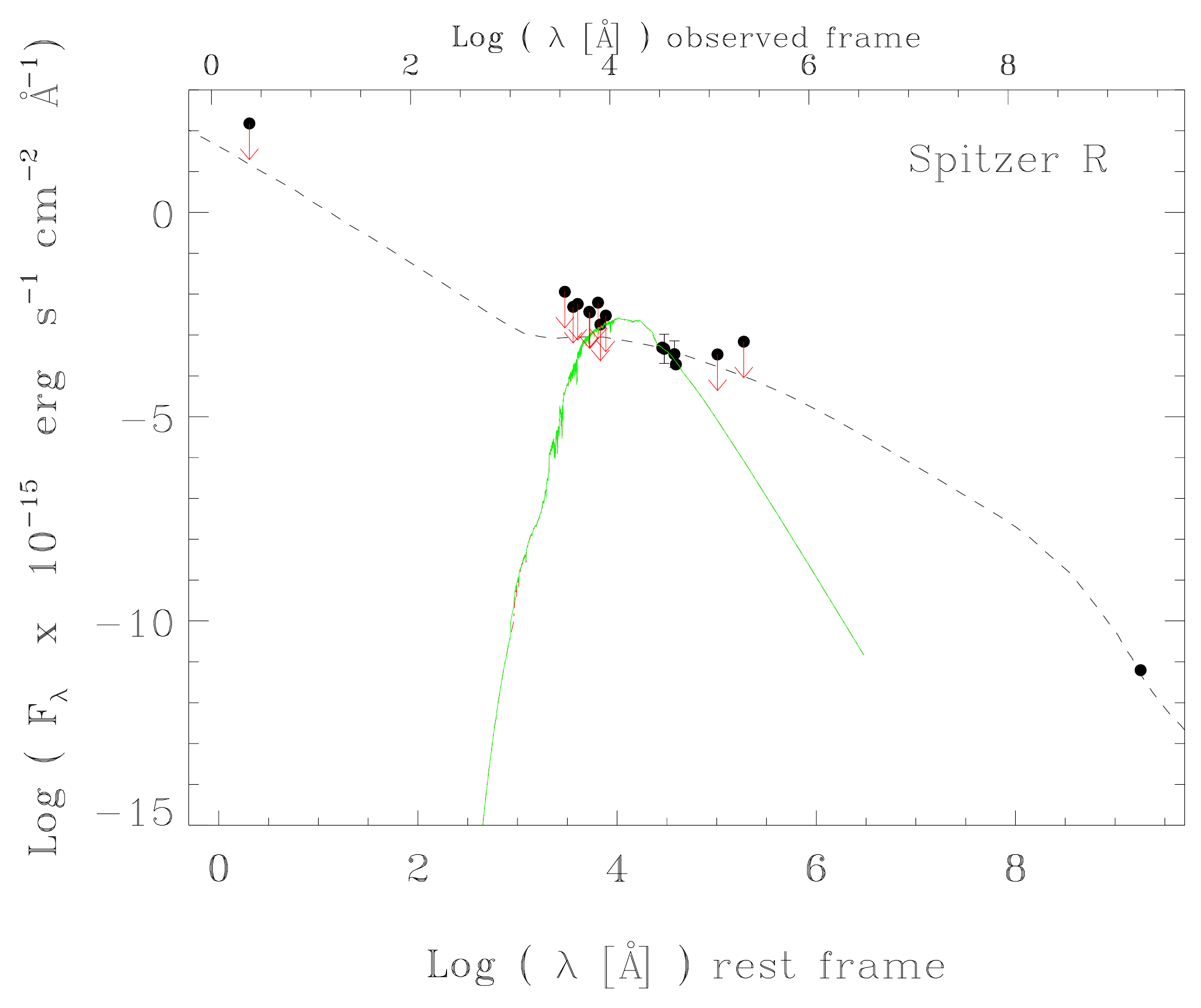}}\\
\caption{Optical-infrared SEDs of the multi-band associations of the X-ray knots (Table~\ref{tab:XrayData})
  and the VLA radio component. The first X-ray knot is associated with
  \textit{Spitzer} source A (upper left panel), X-ray knot 2 with
  \textit{Spitzer} source B (upper right panel), X-ray knot 3 with
  \textit{Spitzer} source C (lower left panel), and then the radio
  component detected with the VLA as \textit{Spitzer} source R (lower
  right panel). The photometric points are fit with the {\it 2SPD}
  fitting code: the stellar population is the red (dot-dashed) line,
  the dust Black-body components are the light-blue (dashed) lines and
  the total model is the green solid line, which also can include the
  emission line template. The black dashed lines are the
    X-ray-radio SEDs of AGN templates taken from \citet{ruiz10} and
    \citet{chiaberge03} (see Sect.~\ref{sec:SED} for details). For all
    the panels, the wavelengths on the top of the plots correspond to
    observed wavelengths, while those on the bottom are in the rest
    frame.}
\label{fig:Knot1aSED}

\end{figure*}

To improve our understanding of the nature of the X-ray knots and
  the radio component R, we included the radio and X-ray datapoints in
  the SEDs and we plotted with AGN templates taken from literature (see
  Figure~\ref{fig:Knot1aSED}): radio-quiet type-I and type-II
  Seyfert/QSO from \citet{ruiz10} and a radio-loud low-luminosity AGN
  from \citet{chiaberge03}. The X-ray-radio SED of X-ray knot 1
  (Spitzer A) is consistent with the SED of a type-II Seyfert galaxy
  (NGC~4507, N$_{H}$ = 4$\times$10$^{23}$ cm$^{-2}$), while the SEDs
  of other two X-ray knots (Spitzer B and C) can be reproduced by a
  type-I quasar (bolometric luminosity of 10$^{10}$ solar
  luminosities, \citealt{hopkins07}). Not surprisingly, the radio
  source R has a SED smilar to a local radio galaxy, NGC~6251
  \citep{chiaberge03}. These SEDs suggest that a background-AGN nature
  of the X-ray and radio sources is a plausible scenario.

%

\subsection{X-ray Data analysis}
\label{sec:X-rayDataAnalysis}
\subsubsection{X-ray spectral fits} 
We first re-binned the spectra using the \verb'grppha' command in FTOOLS. As there were more than 200 counts in the core spectrum, it was binned with 15 counts per bin for both 2007 datasets and we used $\chi^{2}$ statistics. For all the knots in all epochs, and the 2001 and 2002 data for the core region, there were too few counts to fit with $\chi^{2}$ statistics. Hence, we binned these data with 1 count in each bin and used Cash statistics \citep{Cash} to fit them, which is more appropriate in the low-counts regime. The re-binned spectra (see Section~\ref{sec:XrayData}) were loaded into \textsc{xspec} version 12.8.2 \citep{XSpec} and we fitted each spectrum in the energy range 0.3$-$10.0\,keV for each epoch and CCD chip separately to ascertain whether there was any variability in any of the sources across the six years of data. We found no evidence for variability, although we note that there are a low number of counts in some of the observations, especially in the 2001 and 2002 datasets, which makes a conclusion on the variability difficult.

We chose to fit each of the four epochs (2001, 2002, 2007a, 2007b) separately, fitting the X-ray data for each chip simultaneously due to the different responses of the PN, MOS1 and MOS2 detectors. 
We used a cross-calibration weighting factor in the spectral fitting process between the PN, MOS1 and MOS2 detectors. We allowed the cross-calibration constant between the instruments to take into account differences in absolute flux calibration and as such all fluxes given are relative to the PN detector. We fitted all the X-ray knots and the core with the same X-ray spectral models as F17 and compared their results.

\subsubsection{Nuclear region spectral fits}
F17 found that the best fit model (named MEPL) was an absorbed power-law with an additional thermal component due to a collisionally ionised plasma. We give the best fit parameters of the MEPL model from our X-ray spectral analysis in Table~\ref{tab:SpectralFitting} for the 2007 datasets. The X-ray fit parameters (e.g., photon index, see Table~\ref{tab:SpectralFitting}) ascertained from the two 2007 datasets agree with one another within the uncertainties. The parameters for the 2001 and 2002 datasets also agree with the 2007 epochs within the uncertainties, despite having much lower count rates. Therefore, the nuclear region did not vary over the four epochs. As the data for the 2001 and 2002 epochs did not have enough counts in the MOS detectors to use $\chi^{2}$ statistics, we only give the values for the two 2007 datasets fit separately in Table~\ref{tab:SpectralFitting}. We note that the photon index obtained in our analysis is slightly lower than that obtained in F17, likely due to the wider band 0.3$-$10.0\,keV band used for fitting.

\begin{table}
	\centering
	\cprotect\caption{Spectral fits of the nuclear region of X-ray emission in NGC\,6217 for the 2007a and 2007b epochs. The models are outlined in Section~\ref{sec:X-rayDataAnalysis}. $\Gamma$ refers to the photon index of a power law fit, with a respective normalisation of the photon index ``Photon Index, norm". kT and norm refer to the temperature and normalisation of the thermal component. 
The reduced ${\chi}^2$ statistic (${\chi}^2$/d.o.f.) is given for the core, where there are sufficient counts to model the data. The total flux was calculated using the \textsc{xspec} \verb'flux' command, and is given for all models and datasets in the 0.3-10.0\,keV energy range, as well as those corrected for absorption. All fluxes are given in the units of erg\,cm$^{-2}$\,s$^{-1}$. All errors are to 90 \textit{per cent} confidence level.}
	\label{tab:SpectralFitting}
	\begin{tabular}{lccccccccccccr} 
		\hline\hline
		Parameter  & 2007a & 2007b\\
		(1) & (2) & (3)\\
		\hline
		Constant (MOS 1) & 1.03$_{-0.08} ^{+0.09}$ & 0.98$_{-0.07} ^{+0.08}$ \\
		\\[-0.8em]
		Constant (MOS 2) & 1.02$_{-0.08} ^{+0.09}$ & 1.04$_{-0.08} ^{+0.08}$ \\
		\\[-0.8em]
		\hline
		Pho. Index, $\Gamma$ & 2.37$_{-0.18} ^{+0.17}$ & 2.29$_{-0.14} ^{+0.15}$\\
		\\[-0.8em]
		Pho. Index, norm. ($\times$10$^{-5}$)& 2.36$_{-0.34} ^{+0.35}$ & 2.47$_{-0.31} ^{+0.32}$ \\
		\\[-0.8em]
		kT, keV & 0.78$_{-0.02} ^{+0.03}$ & 0.74$_{-0.02} ^{+0.03}$  \\
		\\[-0.8em]
		norm ($\times$10$^{-5}$) & 5.46$_{-0.42} ^{+0.43}$ & 5.32$_{-0.36} ^{+0.37}$ \\
		\\[-0.8em]
		${\chi}^2$/dof & 98/87 & 121/100 \\
		Reduced ${\chi}^2$ & 1.12 & 1.21 \\
		\\[-0.8em]
		\hline
		Total Flux 0.3-10.0\,keV ($\times$10$^{-13}$)& 2.18$_{-0.09} ^{+0.13}$ & 2.20$_{-0.09} ^{+0.12}$ \\
		\\[-0.8em]
		Unabs. Flux 0.3-10.0\,keV ($\times$10$^{-13}$)& 2.55$_{-0.11} ^{+0.14}$ & 2.62$_{-0.13} ^{+0.14}$ \\
		\\[-0.8em]
		\hline
		Total Lum. 0.3-10.0\,keV ($\times$10$^{40}$)& 1.05$_{-0.04} ^{+0.07}$ & 1.06$_{-0.04} ^{+0.06}$ \\
		\\[-0.8em]
		Unabs. Lum. 0.3-10.0\,keV ($\times$10$^{40}$)& 1.23$_{-0.06} ^{+0.09}$ & 1.27$_{-0.06} ^{+0.07}$ \\
		\\[-0.8em]
		\hline		
		\end{tabular}

\end{table}

\subsubsection{Spectral fits for the X-ray knots} 

We employed the same simultaneous fitting method outlined in Section~\ref{sec:X-rayDataAnalysis} for all of the knots separately. We fit an absorbed power-law (as in F17) for the knots and we found that the parameters (photon index, $\Gamma$ and power-law normalisation) agreed with one another between observations, for each knot separately. We used the redshifts obtained from Section~\ref{sec:SED} as well as the redshift for NGC\,6217 \citep[z=0.0045, d=20.1\,Mpc][]{CabreraLavers2004} for the spectral analysis of all the knots, but found no difference in the fit parameters or validity of the fits. 

Although our spectral fits agree with F17 to within the uncertainties, the errors on the fit parameters are large, especially in the 2001/2002 datasets, mainly due to the low number of counts. Overall, in the full spectral range fitted (0.3-10\,keV), there were 291$\pm$28 counts in knot 1, 145$\pm$12 counts for knot 2 and 181$\pm$35 for knot 3. Above 3\,keV, there are 93$\pm$10 counts in knot 1, 37$\pm$9 counts in knot 2 and 28$\pm$6 for knot 3. In addition, we note that for knots 1, 2 and 3, we find photon index values for the 2007 data (1.59$^{+0.27}_{-0.25}$, 1.38$^{+0.45}_{-0.37}$ and 1.54$^{+0.54}_{-0.46}$ respectively) agree within the uncertainties to the values found in F17, but our absolute values are slightly smaller than those in F17 (1.74, 1.71 and 1.70 for each knot respectively). We attribute this discrepancy to using a wider spectral energy range in our procedure, although we note that we get similar results when using the same spectral range of F17. However, we do not attempt to combine all of the data into one dataset for each knot, nor do we attempt a fit of all epoch data for all knots as the photon statistics for each knot is so poor that any approach might turn in similar fit results because of the large uncertainties.

If we assume that the sources are all at the same distance as NGC\,6217, the unabsorbed X-ray luminosities in the 0.3-10.0\,keV energy band for knots 1, 2 and 3 are 3.93$^{+1.13}_{-0.95} \times$10$^{39}$, 2.32$^{+1.80}_{-1.29} \times$10$^{39}$ and 2.37$^{+1.65}_{-1.13} \times$10$^{39}$ erg s$^{-1}$ respectively. However, if we consider the distances found in Section~\ref{sec:SED}, we arrive at values of 7.58$^{+2.18}_{-1.83}\times$10$^{43}$, 1.73$^{+1.35}_{-0.96}\times$10$^{44}$ and 4.13$^{+2.88}_{-1.96}\times$10$^{43}$ erg s$^{-1}$ for knots 1 to 3, respectively.

\section{Discussion}

\subsection{The nuclear region of NGC\,6217}
\label{sec:TheNucleus}
As noted previously, the optical classification of the nucleus is debated in the literature (see Section~\ref{sec:Introduction}) whether its optical spectrum is dominated by an AGN or SF. We investigate two main possibilities: twin-lobed radio emission caused by a LLAGN or a compact star forming region in the centre of NGC\,6217. We also consider whether a combination of both LLAGN activity and SF can explain the observed multi-wavelength emission.

\subsubsection{A LLAGN origin}

As explained in Section~\ref{sec:RadioData}, there are two radio lobe-like structures extending $\sim$3$\arcsec$ north-west and south-east of the optical nucleus. Furthermore, the southern lobe-like feature is more prominent than the northern one, with complex structures and a bright hot-spot. No core has been detected in the radio images. Morphologically, the twin-lobed radio structure of NGC\,6217 is similar to an expanding radio cocoon, consistent with compact steep spectrum (CSS) or gigahertz-peaked (GPS) sources \citep{Orienti2016}. However, CSS and GPS sources typically show two bright hot-spots but we only observe one in NGC\,6217.

The radio structures do not unambiguously match with the line emitting region in the H$\alpha$ \textit{HST} image. However, assuming an astrometric error of 0$\farcs$3 of the \textit{HST} image, we note there exists a small faint region of H$\alpha$ emission, approximately located at the centre of the total radio structure. This optical center likely corresponds to the centre of the galaxy and the location of the AGN. 

By considering the radio and X-ray properties of NGC\,6217, we discuss whether a LLAGN is present at the center of the galaxy. With regard to the X-ray spectra, the photon index ($\Gamma$=2.3) is a little higher than the range reported in the literature for typical AGN \citep[$\Gamma \sim$ 1.5--2.1,][]{NandraPounds, Reeves2000, Piconcelli2005, Page2005, Ishibashi}, although not completely inconsistent with some of the sources reported in those publications. Furthermore, soft X-rays from the inner jet \citep[e.g. ][]{Hardcastle2007} may increase the photon index. Additional, unresolved, nuclear components like an ultra-luminous X-ray sources (ULX) or X-ray binaries (XRB) \citep[e.g. ][]{Stobbart2006} may also add additional soft photons to the nuclear X-ray spectrum. The low resolution of the X-ray data, which encompasses most of the optical extent of the galaxy, makes it impossible to conclude definitively whether an additional nuclear component is adding to the X-ray spectrum. Regarding the radio emission, similar edge-brightened radio structures have been observed with the e-MERLIN array in nearby galaxies hosting LLAGN (see LeMMINGs, \citealt{BaldiLeMMINGs}). In addition, the orientation of the radio lobe structures along the major axis of the galaxy strengthens the AGN scenario, since starburst super-winds tend to produce radio bubbles aligned with the minor axis \citep{Ulvestad1984,Baum1993,Colbert1996,Kinney2000,Gallimore2006,Robitaille2007,Irwin2017}. Therefore, assuming that the radio structure is powered by the central BH, two possible classes of LLAGN can be considered: LINERs and Seyferts.

\begin{figure}
\begin{centering}
	\includegraphics[width=0.45\textwidth]{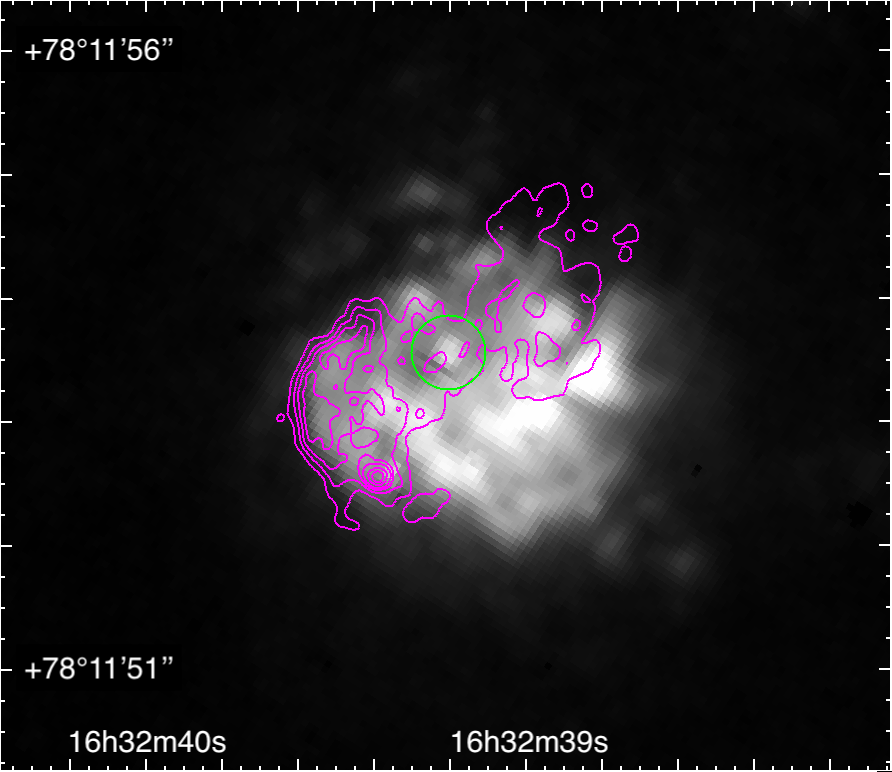}
    \caption{The \textit{HST} H$\alpha$ image of the central 7 $\arcsec$ $\times$ 7 $\arcsec$, corresponding to 680 $\times$ 680 pc, nuclear region in NGC\,6217, overlaid with the radio contours from the combined e-MERLIN/VLA data in magenta. The radio contours are set at 50, 100, 150, 200, 300, 400, 500 $\times$ 10$^{-6}$Jy beam$^{-1}$. The green circle is centred on the hottest pixel in the H$\alpha$ image and the radius corresponds to the typical \textit{HST} offset of 0$\farcs$3. North is up and east is to the left.}
    \label{fig:RadioHAlphaOverlay}
\end{centering}
\end{figure}

LINERs typically show more core-brightened radio morphologies than NGC\,6217, but several cases of parsec-scale double-lobed jetted structures associated with LINERs are present in the LeMMINGs survey. The multi-band luminosities of the nucleus lead to a LINER scenario. In fact, the comparison of the [O~III] line luminosity, measured from the Palomar survey, 1.78 $\times$ 10$^{39}$ erg s$^{-1}$ with the 2-10\,keV unabsorbed nuclear luminosity, set NGC\,6217 in a region occupied mostly by LINER-like sources in the X-ray -- [O~III] diagram (\citealt{Hardcastle2009,Torresi2018}). Furthermore, our e-MERLIN radio observations yield an upper limit of the 1.51\,GHz luminosity of the core of 1.91$\times$10$^{35}$ erg s$^{-1}$, which when compared to the L$_{\rm core}$ -- L$_{\rm [O~III]}$ relation of \cite[their Fig.~6]{BaldiLeMMINGs}, falls in the region of the LINERs. Another argument in favour of the LINER scenario is the location of this target in the fundamental plane of BH activity in the optical band \cite[see][]{Saikia,Saikia2018} within the region covered by LINERs \cite[their Fig.~8]{BaldiLeMMINGs}. Furthermore, F17 extrapolated the NVSS flux for NGC\,6217 to 151\,MHz and found it consistent with a LINER based on radio, infra-red and X-ray properties \citep{Mingo2016}. In addition, F17 calculated the logarithm of the Eddington ratio for NGC\,6217 as -4.09, consistent with a low-level accreting AGN  such as LINERs.

Seyfert galaxies typically show large-scale AGN-driven outflows \citep{Colbert1996}, which appear in radio as double-lobed and less diffuse than radio structures of LINERs \citep{BaldiLeMMINGs}. Similar morphologies have been observed in the Narrow Line Seyfert 1 galaxy Mrk\,783 \citep{Congiu2017}, which has a similar southern edge-brightened lobe and hot spot seen in NGC\,6217. There have been a number of other investigations into Seyfert nuclei showing similar pc--kpc `bubble-like' structures, attributed either to AGN activity, SF or a combination of the two \citep{Wilson1983, Elmouttie1998, kukula99, Hota2006, Kharb2006, Kharb2010, Nyland2017, Ramirez2018}.

Another possible scenario is a 'relic' radio source, supported by the lack of a clear core and the diffuse relaxed bubble morphology. A similar conclusion of a turned-off AGN has been drawn for Mrk\,783 by \cite{Congiu2017}, based on the steep-spectral index of the lobe. To explore this possibility, we calibrated archival VLBA data on NGC\,6217 at 8\,GHz using the \textsc{AIPS} VLBA pipeline \verb'VLBARUN'. The VLBA data show no detection of NGC\,6217 at the 5-sigma level (rms noise 3$\times$10$^{-4}$Jy/beam). These data could either suggest that the core is extremely weak, or the AGN is currently inactive, or that the radio core has an inverted spectral index with a turnover frequency greater than 8\,GHz, similar to that seen in NGC\,7674 \citep{Kharb2017}.

\subsubsection{A star formation origin}
Fig.~\ref{fig:RadioHAlphaOverlay} shows a large amount of nuclear H$\alpha$ emission but not aligned with the radio emission, possibly caused by diffuse SF. In fact, the optical \textit{HST} ACS F814W image of NGC\,6217 shows the presence of strong SF in the galaxy and across the spiral arms. In addition, an inner stellar ring with radius $\sim$1.3$\arcsec$, with a dust spiral roughly at the radius of this inner stellar ring, is present. Certainly, there is SF that is contributing to the observed multi-wavelength nuclear properties of NGC\,6217. Therefore, all these features may suggest a SF related origin of the radio structure of NGC\,6217 observed with e-MERLIN. 

\begin{figure*}
	\begin{center}
		\includegraphics[width=\textwidth]{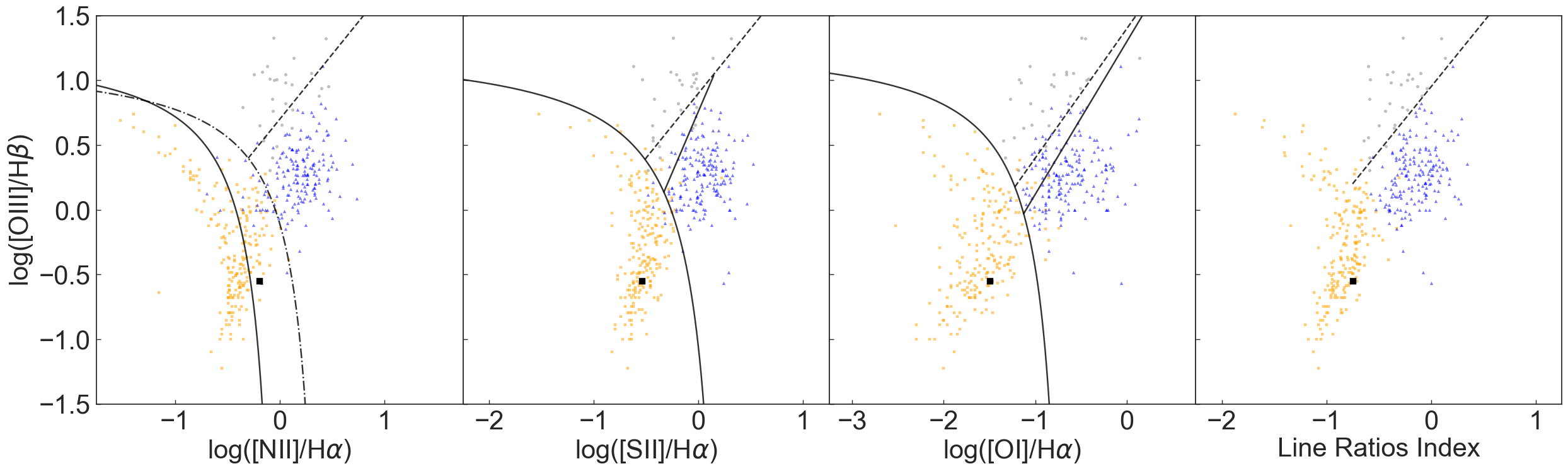}
    \caption{BPT emission line diagnostic diagrams for NGC\,6217. Going from left to right, the panels show the log[O~III]/H$\beta$ ratio against (panel 1) log[N~II]/H$\alpha$, (2) log[S~II]/H$\alpha$, (3) log[O~I]/H$\alpha$ and (4) Line Ratios Index. The position of NGC\,6217 in the diagrams is shown by the black squares. The black lines are the \citet{Kewley2001} diagnostic lines, the dot-dashed line in the first panel shows the \citet{Kewley06} maximum starburst line and the dashed lines show the \citet{Buttiglione2010} classification between LINERs and Seyferts. The BPT diagram for the whole Palomar sample \citep{Ho97a} is shown in the background, with H~II galaxies in orange, LINERs in blue and Seyferts in grey.}
    		\label{fig:BPTDiagram}
	\end{center}
\end{figure*}

The ratios of emission lines taken from the optical spectrum provided by the Palomar survey \citep{Ho97a,Ho97b} ([N~II], [S~II], [O~III] and [O~I] forbidden lines and the H$\alpha$ and H$\beta$ lines) are used to diagnose the nature of gas ionisation in the so called BPT diagrams \citep{BaldwinPhillipsTerlevich81,Kewley06,Buttiglione2010}. NGC\,6217 falls into the region of star forming galaxies in this diagram (Fig.~\ref{fig:BPTDiagram}), consistent with Palomar survey classification \citep{Ho97b}, despite the updated separation between AGN- and SF-dominated area in the BPT diagrams introduced by \citet{Kewley06}. However, it is worth mentioning that the Palomar optical spectra were obtained with a slit of effective aperture 2$-$4$\arcsec$ in size, which would include most of the SF visible in the \textit{HST} image for our object. Therefore, it is also possible that a LLAGN is still hidden in the very centre, but not contributing significantly to the line emission.

Parsec-scale radio bubbles have been observed in local star forming galaxies \citep[e.g. see the LeMMINGs survey][]{BaldiLeMMINGs}. These lobed structures can be driven by super-wind bubbles inflated by supernovae (SNe) \citep{Pacholczyk1976, Heesen2009}. We computed the energy budget of the lobes assuming that they could be driven by a nuclear starburst using a similar process as described in \cite{Kharb2006} for Mrk 6, a local galaxy which a radio structure similar to NGC\,6217. This methodology involves computing the time for a lobe to expand to the observed physical size and comparing this to time taken for SNe to explain the observed emission, assuming adiabatic expansion. We calculate the volume of each lobe assuming the lobes are spherical with a radius (r $\sim$0.3\,kpc). The pressure within the ISM is $\sim$ 10$^{10}$ dynes cm$^{-2}$ and the energy required is therefore 4$PV$ $\approx$ 3.2 $\times$ 10$^{53}$ erg. \cite{Kharb2006} postulate an average SNe energy of 10$^{51}$ erg, which if applied here requires 3.2 $\times$ 10$^2$ SNe in this nuclear region. Assuming wind speeds of 10$^3$\,km\,s$^{-1}$, it would take $\sim$ 3 $\times$ 10$^5$ years to inflate the `bubble' to its present size, requiring a SNe rate of $\sim$ 1.1 $\times$ 10$^{-3}$ per year. This value for the SNe rate is similar to that calculated for M~82 for a nuclear region of a similar size to that shown here for NGC\,6217 \citep{BaldiLeMMINGs}. Hence, this calculation energetically shows that the SF scenario for the radio emission of NGC\,6217 is plausible.

Another argument in favour of the SF scenario is the low brightness temperature T$_{B}$ observed in the NGC\,6217. T$_{B}$ is a metric that is often used to discriminate compact, star forming H~II regions from AGN as H~II regions tend to be limited to T$_{B}$ $\la$ 10$^{5}$K \citep{Condon92}. For objects above this threshold, additional emission from relativistic electrons from the AGN is required to explain the high brightness temperature. The low T$_{B}$ measured for our target suggests that SF can account for the low-brightness diffuse emission detected in the e-MERLIN image. However, as pointed out by \citet{BaldiLeMMINGs}, since brightness temperatures depend on the flux densities and the sizes, at e-MERLIN resolution and sensitivity only sources brighter than 5 mJy beam$^{-1}$ can yield brightness temperatures greater than 10$^{6}$ K. Therefore, even considering the brightest unresolved component in NGC\,6217, which is the southern hot spot with 0.45\,mJy, T$_{B}$ is far below the separation threshold between AGN and SF. Therefore, the low T$_{B}$ cannot provide a conclusive answer to the nature of the the faint extended radio emission in NGC\,6217. 

\subsubsection{A LLAGN and SF origin}

Due to the large amount of contrary evidence, it is difficult to conclude with certainty whether the observed radio emission is due to SF or a LLAGN. Another possibility for the observed multi-band emission, is that both an LLAGN and SF contribute to different observed properties. This interpretation is supported by the large ongoing SF visible in the \textit{HST} images and the radio and X-ray properties, more consistent with a phenomenon of accretion onto a supermassive BH. We conclude that the nuclear region of NGC\,6217 probably contains a LINER-type LLAGN, similar to some of the H~II galaxies observed in the LeMMINGs `shallow' sample \citep{BaldiLeMMINGs}. However, the contribution from second nuclear component like X-ray binaries cannot be ruled out given the data resolution. Hence, we agree with the conclusions of F17 regarding the combined AGN and SF origin of the nuclear emission. Higher resolution (e.g. \textit{Chandra}) X-ray data is required to properly understand the nuclear properties of this source.

\subsection{The putative X-ray Jet of NGC\,6217}
\label{sec:TheXrayJet}

A detailed analysis of the X-ray data in NGC\,6217 is presented by F17, who consider \textit{XMM-Newton} observations. In addition to emission from the galaxy itself, they note three X-ray emission knots in a line along a P.A. of $\sim$225$^{\circ}$ (Fig~\ref{fig:HSTRadioXMMSpitzer}), oriented towards the galaxy, with the most distant being $\sim$15\,kpc from the nucleus. They find no optical associations in the XMM-Optical Monitor (OM) images or the SDSS data for knot 1. For knot 2, F17 find one associated source, at 1.6$\arcsec$ away from the X-ray position in the OM $u$ filter and a further source in the SDSS $r$-band at 0.9$\arcsec$ away, but conclude that these sources are likely unrelated stars. For knot 3, F17 found an association with an SDSS source 1.2$\arcsec$ away but conclude that it is likely from a passive spiral galaxy.

F17 note that radio emission from the galaxy in low resolution NVSS images shows a low-brightness 2$\arcmin$ elongation along the line of the three X-ray components. They consider various scenarios to explain the X-ray and radio emission components, including background sources, but eventually favour a jet origin. They do however, note that higher resolution radio observations are required to confirm at least the radio extension as coming from a jet. Here, in the light of such higher resolution radio observations, we question the jet origin of the radio and X-ray emission. 

We first consider the radio emission. From our e-MERLIN and VLA observations we can clearly see emission on the scale of a few arcseconds associated with the galaxy itself. In addition, three very compact (sub-arcsecond) sources are observed, one of which lies exactly along the putative radio extension shown in the NVSS images but not associated with the X-ray knots. We explore the nature of this radio component in Section 4.3. It is unlikely that a radio jet structure exists in the intermediate scales between the resolution of the VLA in A Array at 1.51\,GHz (beam size $\sim$1$\arcsec$) and the NVSS survey (beam size $\sim$45$\arcsec$) as archival VLA data\footnote{Images obtained from the VLA Imaging Pilot https://archive.nrao.edu/nvas/} in C configuration and at 1.51\,GHz (beam size $\sim$16$\arcsec$) show only an unresolved source and no extension in any direction.  Moreover, \cite{vanDrielButa} show no extension towards the X-ray jet in H~I images of NGC\,6217 at a resolution of 20$\arcsec$. Thus, from this new high resolution data and archival datasets, it is possible to exclude the presence of a large-scale radio jet associated with the X-ray jet in NGC 6217. The lack of radio jet leaves a gap in our knowledge of the aligned X-ray knots and motivates new high resolutions X-ray observations with \textit{Chandra}. We discuss the scenario where all of the X-ray knots are possibly background sources in the next two sections.

The interpretation of a lack of a radio emitting jet is supported by the global properties of NGC\,6217. In fact, there are no examples in the literature of sources similar to our target, i.e. low-luminosity radio-quiet AGN with a small BH mass (2.0 $\times$ 10$^6$ M$_{\odot}$), which can launch radio jets that are also visible in the X-rays. This is a restrictive prerogative of powerful radio-loud AGN, which show large kpc-scale collimated jets and massive BH ($>$10$^{8}$ M$_{\odot}$). There are some rare cases of X-ray jets observed without any associated radio jet \citep[see for example][]{Simionescu2016,Meyer2017,Hlavacek2017}, but we note that these jets are found in very luminous galaxies with powerful AGN also, unlike NGC\,6217.

\subsubsection{Chance Alignment of the X-Ray knots}

We now consider the likelihood of the X-ray knots appearing to align by chance. First, we ascertain whether the \textit{Spitzer} sources are genuinely related to the X-ray knots. We computed the number of 3$\sigma$ \textit{Spitzer} detections that are expected in one of the circular 21$\arcsec$ X-ray extraction regions, from the \textit{Spitzer} catalogue (e.g. the red squares in Fig~\ref{fig:HSTRadioXMMSpitzer}). We found 148 \textit{Spitzer} detections in the 5.2$\arcmin$ $\times$ 5.2$\arcmin$ \textit{Spitzer} CCD field-of-view, corresponding to $\sim$2 sources per 21$\arcsec$ radius spectral extraction region. We find two 3$\sigma$ detections in the spectral extraction region of knots 1 and 2, and one detection in the spectral extraction region for knot 3, although a second object is very close to said region (see Fig~\ref{fig:HSTRadioXMMSpitzer}). Hence the number of \textit{Spitzer} objects in each extraction region is consistent with the number expected from this simple analysis.

Further to the analysis of the co-spatiality between \textit{Spitzer} sources and the X-ray knots, we calculated the likelihood of the three X-ray knots forming a line from the centre of NGC\,6217 by chance. To perform this statistical analysis we used the integral source counts for \textit{XMM-Newton} \citep{Ranalli2013}. First, we calculated the number of sources we would expect in a given radius from the nucleus to knot 1. We then found the number of sources in the radius from knot 1 to knot 2 and multiplied this value by the angle subtended by knot 2 on knot 1. We performed this step again, but for knots 2 and 3. The product of these three values gives us a small probability of chance alignment of 1.37$\times$10$^{-4}$. However, we also consider the sources that both the NVSS and \textit{XMM-Newton} have observed to estimate the likelihood of seeing such an alignment present in NGC\,6217. As an example, we use the NGC catalogue, which has 7840 galaxies present. Of these, 7074 sources have been observed by the NVSS. Of these 7074 galaxies, 1455 have \textit{XMM-Newton} data in the archive. Therefore, the likelihood of seeing such a chance alignment in the \textit{XMM-Newton} data which is unresolved by the NVSS data is 1.37$\times$10$^{-4}$ multiplied by 1455, yielding a value of $\sim$0.2. 

To test the possibility that three X-ray sources could align on the sky within 3$\arcmin$, purely from random chance, we ran simulations. Within a 3$\arcmin$ box, we placed three points in uniformly random positions many times. For each iteration, we estimated whether all three sources were aligned with the central source at the origin. To perform this step, we traced out an area from the origin given by the width of the central source (30$\arcsec$ in the case of NGC\,6217) projected along the line of the closest source to the origin, out to the edge of the  3$\arcmin$ box. If all of the three points were found in this error region, we considered them aligned.
 We then ran enough iterations that additional runs resulted in negligible statistical improvement. We found that the probability of three knots aligning is $\sim$\,0.3 \textit{per cent}, which is fully consistent with our probabilistic estimate from the \textit{XMM}-Newton source counts.

A more detailed, simulation-based, analysis would be needed to put uncertainties on both our estimated probability values, but variation by greater than a factor of 2, either way, is unlikely. Thus we cannot rule out the null hypothesis that the alignment is just a chance coincidence. Furthermore, these values should be treated as an upper limit since we do not consider the probability that the spectra, fluxes and hardness ratios (see F17) are similar among the knots. 

\subsubsection{The nature of the X-ray knots}

We now consider the three X-ray knots in turn and attempt to explain plausible origins for these X-ray knots, given the lack of co-spatial radio emission and multi-wavelength associations. From the extracted optical-IR sources associated with the X-ray knots within the 21$\arcsec$ X-ray extraction region, we built up their SEDs (see Section~\ref{sec:SED} and Table~\ref{tab:SEDFits}). The genuine physical relation between the X-ray and optical-IR associations goes beyond our reasonable attempt to investigate the nature of these knots. Therefore, caution needs to be taken when considering the physical properties derived for these sources.

Concerning the multi-band identifications of knot 1, two \textit{Spitzer} sources fall within the 21$\arcsec$ X-ray extraction region. The SED fit for the source nearest to the X-ray knot is limited by the lack of data-points. However, we calculate a photometric redshift of 0.48$^{+0.25}_{-0.12}$ with a bolometric luminosity of 7.9$\times$10$^{43}$ erg s$^{-1}$ for this source.
We also extracted the SED of the other \textit{Spitzer} source found in the extraction region of knot 1.  This source is detected in SDSS \textit{r},\textit{i} and \textit{z} as well as all the IR bands and we found a photometric redshift of 0.49$^{+0.16}_{-0.13}$, consistent with the former \textit{Spitzer} source. Given these similar redshifts and their proximity to one another, their separation can be estimated to be of order of 100\,kpc and hence may be part of a cluster of galaxies. 

Similar to knot 1, two \textit{Spitzer} sources fall within the X-ray spectral extraction radius of knot 2, but unlike knot 1, one of the two \textit{Spitzer} sources well matches the centroid of X-ray knot 2 within 0.5$\arcsec$. This \textit{Spitzer} source also has SDSS and \textit{WISE} detections, as well as a \textit{HST} detection. This co-spatiality increases the probability that the X-ray knot and the optical-IR associations are physically related. It should be noted that the optical/IR detections described here are not related to the background star mentioned in F17, 1.6$\arcsec$ away from the knot. The optical-IR SED fit requires the inclusion of the spectrum of a Seyfert-like AGN with strong emission lines which constrains its photo-z at 0.70$^{+0.03}_{-0.02}$ with a bolometric luminosity of 8.3$\times$10$^{43}$ erg s$^{-1}$. 

Finally, X-ray knot 3 has a complicated X-ray morphology, with two distinct X-ray blobs in the X-ray extraction region. There is a \textit{Spitzer} source coincident with the western X-ray component in this knot (see Fig.~\ref{fig:HSTRadioXMMSpitzer}), which is also coincident with the extended SDSS source discussed by F17 and a \textit{HST} source that F17 attributed to an unrelated galaxy. 
Assuming that the optical-IR associations are related to the emission of the knot 3, we modelled the SED with a young and old stellar population with a photo-z of 0.46$^{+0.19}_{-0.06}$ and a bolometric luminosity of 9.3$\times$10$^{43}$ erg s$^{-1}$. This tentative multi-band identification of the X-ray emission from knot 3 makes the nature of this knot as a background source questionable.

Although caution on the interpretation of these knots is needed, the multi-band associations found for the X-ray knots may suggest an extragalactic background origin for these components. This interpretation is supported by the large inferred X-ray and bolometric luminosities (see Table~\ref{tab:SEDFits}), ($\sim$10$^{43}$ erg s$^{-1}$, assuming the photo-z are correct) and X-ray photon indices consistent with an AGN origin \citep{NandraPounds, Reeves2000, Piconcelli2005, Page2005, Ishibashi}. Specifically, the ratio between the limits on the radio and X-ray luminosities are similar to those found in Seyfert galaxies and other low-luminosity AGN \cite[e.g.][]{panessa07,laor08}. However, in the case that the sources belong to the host galaxy, NGC\,6217, all of the knots would have luminosities ($\sim$10$^{39}$ erg s$^{-1}$), which could be consistent with different types of source, such as X-ray binaries or being part of a jet. 
 Therefore, we cannot put firm conclusions on the nature of any of these multi-wavelength associations to the X-ray knots. 
 Further high resolution data with \textit{Chandra} will greatly improve our understanding of these sources, by providing higher angular and spectral resolution data which will allow for true X-ray counterparts to be found for each of the associations.

\subsection{Radio Emission from secondary sources in NGC\,6217}

For the sources that were within the optical extent of the galaxy (panels B and C in Fig~\ref{fig:HSTRadioXMMSpitzer}), we did not attempt to extract their SEDs or look for X-ray emission because emission from the galaxy would confuse the SED fits and the \textit{XMM-Newton} data is not of sufficient resolution to identify these objects separately. Both sources have e-MERLIN flux densities of $\la$5\,mJy at a size of $0\farcs15 \times 0\farcs15$, and therefore have brightness temperatures too low to claim a clear AGN activity. However, these do not align with the star forming regions found by \cite{Gusev2012} and \cite{Gusev2016} in NGC\,6217. The source found in the spiral arms (Panel~B in Figure~\ref{fig:HSTRadioXMMSpitzer}) shows an extended structure and is co-incident with a 3$\sigma$ \textit{Spitzer} source. Judging from the radio emission alone, these radio sources are possibly either H~II regions or background AGN. Additional high-resolution X-ray data would help in to identify these sources.

The radio source located between the X-ray knots 1 and 2 (Panel D) is associated with a weak \textit{Spitzer} source (R). Modelling the only detected IR part of its SED provides an inconclusive photo-z of $0.19^{+0.25}_{-0.12}$ with a bolometric luminosity of 4.6$\times$10$^{42}$\,erg s$^{-1}$. This source appears compact and unresolved at e-MERLIN and VLA A-Array baseline scales, with similar flux density values and appears unrelated to the X-ray emission of the contiguous knots. The most simple interpretation of this source is a background galaxy associated with a low-luminosity radio AGN, with a radio structure confined in a kpc scale. This scenario is in line with the population of compact radio galaxies \citep[named FR~0s,][]{baldi15,baldi18} found in the local Universe. In addition, it also likely that this radio source contributes to the radio elongation observed in NVSS at larger beam sizes along the putative X-ray jet.

\section{Conclusions}
\label{sec:Conclusions}
We analysed new high resolution radio interferometric data of the barred spiral galaxy NGC\,6217 with the high angular resolution and sensitivity of e-MERLIN. For the first time, we have resolved the nuclear region of NGC\,6217, which consists of two radio lobes projected north-west to south-east. The nuclear morphology is similar to compact steep spectrum (CSS) or gighertz-peaked (GPS) sources \citep{Orienti2016} but shows no core component. Starburst super-winds have been observed to show similar morphologies, but usually oriented along the minor axis of the galaxy \citep{Ulvestad1984,Kinney2000,Gallimore2006}. The \textit{XMM-Newton} X-ray spectra suggest a good fit (${\chi}^2$/d.o.f. $\sim$ 1) for an absorbed power law, with an additional thermal component. The energetics, optical line ratios and radio brightness temperature of the core region cannot rule out SF as the main contributor to the emission, but are also consistent with a LLAGN emission origin. Given the radio morphology, X-ray spectra and lack of alignment with the galaxy minor axis, we conclude that the nuclear region is most likely powered by a LLAGN, probably with a LINER type nucleus, with multi-band contamination from SF or other nuclear X-ray sources. Therefore, the 400-pc twin-lobed structure of NGC\,6217 constitutes one of the smallest radio structures observed in the nuclei of nearby LLAGN. To confirm the emission mechanism, further high resolution X-ray data (e.g. \textit{Chandra}) and optical (photometry and spectroscopy from \textit{HST}) data are needed.

The `putative' X-ray jet in NGC\,6217 is not coincident with any radio emission observed with e-MERLIN or the VLA at 1.51\,GHz. Furthermore, the extension of the nuclear radio bubble is oriented almost perpendicular to the putative X-ray jet. We therefore reconsidered the background source origin of the X-ray knots as we found multi-band sources co-incident within the X-ray spectral extraction regions. By extracting photometry of all these sources in the optical/IR bands, we performed spectral energy distribution fits and found that all of the X-ray `knots' were plausibly consistent with higher redshift ($\gtrsim$ 0.4) sources such as background AGN and galaxies and therefore likely not belonging to NGC\,6217. However, the sparse multi-band coverage of the SEDs of the X-ray knots does not ensure the correctness of the physical parameters inferred for these knots and therefore we cannot come to a definite conclusion on the nature of these sources as they are intrinsically faint. Furthermore, due to the high source density of \textit{Spitzer} sources expected in the spectral extraction regions, it is possible that these IR sources are chance associations with the X-ray `knots'. We calculated that the likelihood of the `knots' aligning in such a way to be P=0.2 based on \textit{XMM-Newton} integral source counts. By simulating aligned sources in random positions in a square field equal to the length of the X-ray `knot' structure, we found that such alignments occur in 0.3 \textit{per cent} of cases. Thus we cannot rule out the null hypothesis that the alignment is just a chance coincidence. Therefore, these X-ray knots would benefit from higher resolution X-ray data to confirm their optical/IR associations. However, given that there are no radio associations to the X-ray knots, we have shown that the X-ray jet hypothesis is unlikely in this source.

We also investigated the other radio components observed in the new high-resolution images of NGC\,6217. We found a radio source that coincidentally aligned along the P.A. of the putative X-ray jet, but not associated with any of the X-ray emission. We suggest that this may be due to a background radio AGN of redshift $\sim$0.19. The two other sources of radio emission are possibly H~II regions in the galaxy or background AGN, but additional data are required to test this hypothesis.

The study of radio emission in the nuclear regions of nearby LLAGN is important for discriminating SF from AGN activity. Our observations of NGC\,6217 show that such studies can only be undertaken with the unique combination of good angular and spectral resolution provided by e-MERLIN. By combining e-MERLIN data with the VLA, it is possible to probe multiple spatial scales simultaneously. However, understanding how nuclear regions of nearby galaxies are ionised, whether by an LLAGN of SF, is only possible by comparing radio data with high-resolution optical and X-ray data. Furthermore, the LeMMINGs sample allows for such studies to be undertaken in a statistically complete way to compare the different radio morphologies across all nuclear optical classifications. 

\section*{Acknowledgements}

We thank the anonymous referee for useful comments that helped to improve the manuscript. We acknowledge funding from the Mayflower Scholarship from the University of Southampton awarded to David Williams to complete this work. DW was supported by the Oxford Centre for Astrophysical Surveys, which is funded through generous support from the Hintze Family Charitable Foundation. IMcH thanks the Royal Society for the award of a Royal Society Leverhulme Trust Senior Research Fellowship. RDB and IMcH also acknowledge the support of STFC under grant [ST/M001326/1]. A.B. is grateful to the Royal Society and SERB (Science and Engineering Research Board, India) for financial support through the Newton-Bhabha Fund. A.B. is supported by an INSPIRE Faculty grant (DST/INSPIRE/03/2018/001265) by the Department of Science and Technology, Govt. of India. P.B. acknowledges the STFC for financial support. JHK acknowledges financial support from the European Union's Horizon 2020 research and innovation programme under Marie Sk\l{}odowska-Curie grant agreement No 721463 to the SUNDIAL ITN network, from the Spanish Ministry of Economy and Competitiveness (MINECO) under grant number AYA2016-76219-P, from the Fundaci\'{o}n BBVA under its 2017 programme of assistance to scientific research groups, for the project "Using machine-learning techniques to drag galaxies from the noise in deep imaging", and from the Leverhulme Foundation through the award of a Visiting Professorship at LJMU. DMF wishes to acknowledge funding from an STFC Q10 consolidated grant [ST/M001334/1]. EB and JW acknowledge support from the UK's Science and Technology Facilities Council [ST/M503514/1] and [ST/M001008/1], respectively. We thank Dr S\'{e}bastien Comer\'{o}n for his help with the reduction of the HST ACS H$\alpha$ image.

We acknowledge Jodrell Bank Centre for Astrophysics, which is funded by the STFC. e-MERLIN and formerly, MERLIN, is a National Facility operated by the University of Manchester at Jodrell Bank Observatory on behalf of STFC. Some of the observations in this paper were made with the NASA/ESA Hubble Space Telescope, and obtained from the Hubble Legacy Archive, which is a collaboration between the Space Telescope Science Institute (STScI/NASA), the European Space Agency (ST-ECF/ESAC/ESA) and the Canadian Astronomy Data Centre (CADC/NRC/CSA). This publication makes use of data products from the Wide-field Infrared Survey Explorer, which is a joint project of the University of California, Los Angeles, and the Jet Propulsion Laboratory/California Institute of Technology, funded by the National Aeronautics and Space Administration. IRAF is distributed by the National Optical Astronomy Observatory, which is operated by the Association of Universities for Research in Astronomy (AURA) under a cooperative agreement with the National Science Foundation. Funding for the Sloan Digital Sky Survey IV has been provided by the Alfred P. Sloan Foundation, the U.S. Department of Energy Office of Science, and the Participating Institutions. SDSS acknowledges support and resources from the Center for High-Performance Computing at the University of Utah. The SDSS web site is www.sdss.org.

SDSS is managed by the Astrophysical Research Consortium for the Participating Institutions of the SDSS Collaboration including the Brazilian Participation Group, the Carnegie Institution for Science, Carnegie Mellon University, the Chilean Participation Group, the French Participation Group, Harvard-Smithsonian Center for Astrophysics, Instituto de Astrof\'{i}sica de Canarias, The Johns Hopkins University, Kavli Institute for the Physics and Mathematics of the Universe (IPMU) / University of Tokyo, Lawrence Berkeley National Laboratory, Leibniz Institut f\"{u}r Astrophysik Potsdam (AIP), Max-Planck-Institut f\"{u}r Astronomie (MPIA Heidelberg), Max-Planck-Institut f\"{u}r Astrophysik (MPA Garching), Max-Planck-Institut f\"{u}r Extraterrestrische Physik (MPE), National Astronomical Observatories of China, New Mexico State University, New York University, University of Notre Dame, Observat\'{o}rio Nacional / MCTI, The Ohio State University, Pennsylvania State University, Shanghai Astronomical Observatory, United Kingdom Participation Group, Universidad Nacional Aut\'{o}noma de M\'{e}xico, University of Arizona, University of Colorado Boulder, University of Oxford, University of Portsmouth, University of Utah, University of Virginia, University of Washington, University of Wisconsin, Vanderbilt University, and Yale University.

DW would also like to thank Charlotte Angus, Daniel Asmus, Marco Chiaberge, Sam Connolly, Rob Firth, Poshak Gandhi, Javier Mold\'{o}n, Sara Motta, Francesca Panessa, Jack Radcliffe, Anthony Rushton, Mat Smith and Lorenzo Zanisi for useful discussions relevant to this paper.



\bibliographystyle{mnras}

\bibliography{mybibliography.bib}



%
%
%


\bsp	
\label{lastpage}
\end{document}